\documentclass[fleqn,usenatbib]{mnras}

\usepackage{newtxtext,newtxmath}

\usepackage[T1]{fontenc}
\usepackage{ae,aecompl}

\usepackage{graphicx}	
\usepackage{amsmath}	

\usepackage{comment}

\defcitealias{Shrestha_2018}{Paper I}

\title[Bow shock polarization -- dust]{\textit{Polarization simulations of stellar wind bow shock nebulae. \\ II. The case of dust scattering}}

\author[M. Shrestha et al.]{
Manisha Shrestha,$^{1,2}$\thanks{E-mail: manisha.shrestha9@du.edu (DU)}
Hilding R. Neilson,$^{3}$
Jennifer L. Hoffman,$^{1}$
Richard Ignace,$^{4}$
\newauthor Andrew G. Fullard $^{1}$
\\
$^{1}$Department of Physics \& Astronomy, University of Denver, 2112 E Wesley Ave., 80208, US\\
$^{2}$Astrophysics Research Institute, Liverpool John Moores University, 146 Brownlow Hill, Liverpool L3 5RF, UK\\
$^{3}$Department of Astronomy \& Astrophysics, University of Toronto, 50 St. George Street, M5S 3H4, Canada\\
$^{4}$Department of Physics \& Astronomy, East Tennessee State University, Johnson City, TN, 37614, US
}

\date{Accepted XXX. Received YYY; in original form ZZZ}

\pubyear{2020}

\begin{document}
\maketitle

\begin{abstract}
We study the polarization produced by scattering from dust in a bow shock-shaped region of enhanced density surrounding a stellar source, using the Monte Carlo radiative transfer code \textit{SLIP}. Bow shocks are structures formed by the interaction of the winds of fast-moving stars with the interstellar medium. Our previous study focused on the polarization produced in these structures by electron scattering; we showed that polarization is highly dependent on inclination angle and that multiple scattering changes the shape and degree of polarization. In contrast to electron scattering, dust scattering is wavelength-dependent, which changes the polarization behaviour. Here we explore different dust particle sizes and compositions and generate polarized spectral energy distributions for each case. We find that the polarization SED behaviour depends on the dust composition and grain size. Including dust emission leads to polarization changes with temperature at higher optical depth in ways that are sensitive to the orientation of the bow shock. In various scenarios and under certain assumptions, our simulations can constrain the optical depth and dust properties of resolved and unresolved bow shock-shaped scattering regions.Constraints on optical depth can provide estimates of local ISM density for observed bow shocks. We also study the impact of dust grains filling the region between the star and bow shock. We see that as the density of dust between the star and bow shock increases, the resulting polarization is suppressed for all the optical depth regimes.

\end{abstract}




\section{Introduction}
\label{sec:intro}

A stellar wind bow shock nebula is formed by the interaction of a stellar wind with the ambient interstellar medium (ISM) when the relative velocity between the two is supersonic \citep[e.g.,][]{Wilkin_1996}. Because the characteristics of these nebulae depend strongly on the properties of the stellar wind and local ISM, studying them can reveal details of the star's mass loss history and evolution \citep[e.g.,][]{Langer_2012,Kobulnicky_2018}, as well as the structure, evolution, and dust properties of the ISM \citep{Cotera_2001,Ueta_2008,Rauch_2013}.

In a previous study (\citealt{Shrestha_2018}, hereafter Paper I), we carried out a computational investigation of the polarization arising from stellar wind bow shock nebulae when electron scattering is the only polarizing  mechanism. We considered illumination by the star alone and by a ``distributed source," which represents emission from within the bow shock nebula itself. We found that the polarization thus produced is highly dependent on viewing angle, and that multiple scattering modifies the polarization significantly from the analytical predictions for single scattering. In cases involving significant multiple scattering, in addition to a polarization peak near a viewing angle of $90\degr$ predicted by single scattering models, our simulations also produced a second peak at a larger angle. 

As an extension of the work presented in \citetalias{Shrestha_2018}, we expand the study to explore the effects of dust scattering on the polarization behavior of stellar wind bow shock nebulae. For the distance scale of a typical bow shock, dust plays an important role in the density structure of bow shock nebulae around hot and cold massive stars, and as a result many observational studies of these phenomena focus on the various infrared bands ranging from $3.6~\mu$m to $160~\mu$m \citep[e.g,][]{Ueta_2008,Kobulnicky_2016,Jayasinghe_2019}. Consequently, we cannot ignore the role of dust in scattering  light and producing polarization in stellar wind bow shock nebulae. The polarization observed in dusty bow shock nebulae near the Galactic centre has a magnitude as high as a few per cent \citep{Buchholz_2012,Rauch_2013, Shahzamanian_2016,zajacksum_2017}. 

Several authors have previously modeled the polarimetric features arising from dust scattering in bow shock structures. \citet{Buchholz_2012} used analytical calculations, which are applicable only for single scattering at very low optical depths. \cite{Shahzamanian_2016} and \cite{Zajacek_2017} used a sophisticated 3-D Monte Carlo radiative transfer (MCRT) code to simulate the polarization behaviour of a dust-scattering bow shock along with other possible circumstellar structures around the Dusty S-cluster Object (DSO) near the Galactic centre. These studies focused on a particular object and included scattering regions other than the bow shock nebula itself. Our aim is to build on these previous models and create numerical simulations for a generalized bow shock structure that include the effects of multiple scattering, to allow consideration of higher optical depth regions.

This contribution is the second of two papers in which we use the MCRT method to simulate the polarimetric behavior of generalised stellar wind bow shock structures. We obtained our results using the \textit{SLIP} code (``Supernova LIne Polarization"; \citealt{Hoffman_2007}; \citetalias{Shrestha_2018}).
This code is similar to the one used by \cite{Shahzamanian_2016} and \cite{Zajacek_2017}, but our implementation is different, as discussed in \citetalias{Shrestha_2018}. The ultimate goal of our study is to determine how polarization measurements may constrain the properties of the bow shock, which in turn provides constraints on the properties of the interstellar medium and the stellar wind that produces the bow shock. Here we investigate the effects of various input parameters on the resulting polarization behavior, assuming dust is the only scattering mechanism. As in \citetalias{Shrestha_2018}, we will use the term ``bow shock'' in a broad sense, describing not only a physical shock, but also the resulting nebula, or region of enhanced density, surrounding the shock and having the same shape. With respect to the bow shock classification introduced by \citet{Henney_2019},our models represent radiation bow shocks and radiation-supported bow waves, for the cases of low and high optical depth, respectively.

We note that magnetic fields can align dust grains in a way that produces polarization; they also play an important role in both the morphology and the emission properties of a stellar wind bow shock \citep[e.g.,][]{Meyer_2017,Henney_2019b}. However, treating the complex effects of magnetic fields is beyond the scope of this paper. We consider only spherical grains in our simulations, and do not attempt to simulate other magnetic field effects. Alignment of interstellar grains by Galactic magnetic fields can produce interstellar polarization (ISP, initially described by \citealt{Serkowski_1975}), which manifests as an additional component of observed polarization beyond the local effects we consider. We do not simulate ISP in these models because its properties vary strongly with Galactic sightline, and because various methods exist to estimate and remove it from polarization observations \citep[see, e.g.,][]{quirrenbach1997}. We discuss ways to account for ISP in relevant sections below.

This paper is organized as follows. In Section~\ref{method} we discuss the implementation of dust scattering in the \textit{SLIP} code and provide details regarding the dust models we adopted in our simulations. In Section~\ref{comp}, we present results from an analytical model compared with our numerical models, and also compare our dust-scattering results with those of models assuming electron scattering only. In Section~\ref{results} we present and interpret our  model predictions for different emission sources, dust types, and wavelengths, and for both resolved and unresolved cases. In Section~\ref{comparison_observation}, we discuss observational implications and compare the results from our simulations with observational data. Finally, conclusions and future work are presented in Section~\ref{conclusions}.

\section{Methods}\label{method}
We created all simulations presented here with the \textit{SLIP} code (\citealt{Hoffman_2007}; \citetalias{Shrestha_2018}). \textit{SLIP} is a MCRT code \citep{whitney_2011} that tracks photon packets through a three-dimensional spherical polar grid \citep{Whitney_2002}. We assume a $\phi$-symmetric case with 100 radial cells and 101 cells in the polar ($\theta$) direction, linearly spaced in both coordinates.

A finite spherical photon source (1 $R_\odot$) sits at the center of the grid, surrounded by circumstellar material (CSM) composed of dust particles in local thermodynamic equilibrium (LTE; \citealt{Mohamed_2012}). The code does not perform radiative equilibrium calculations to account for heating of the CSM by the central star or shock. Instead we specify a dust temperature $T_d$ at a reference radial distance $r_d$ from the star as an input parameter, and assume the temperature in the rest of the CSM decreases with distance $r$ from the central star. Setting the  bolometric flux emitted by the dust at temperature $T$ equal to the stellar flux at the corresponding distance $r$, we find that the CSM temperature is given by $T=T_d\sqrt\frac{r_d}{r}$. (We take the reference radius $r_d$ to be $R_0$, the bow shock standoff radius; see Eq.~\ref{standoff} below.) This $T$ governs the emission from the dust within the CSM. Although this calculation does not include heating of the dust by the shock itself, it allows a more physical treatment of the temperature than in \citetalias{Shrestha_2018}, where we assumed a constant $T$ throughout the scattering region. We also specify a reference optical depth $\tau_0$ at a convenient arbitrary reference angle, $\theta_0=1.76$ rad $= 95.4\degr$, the same as in \citetalias{Shrestha_2018}. 

\textit{SLIP} emits virtual, initially unpolarized ``photons" from the central star (or other photon source) and tracks them as they travel through the CSM. Each photon's behavior is determined by generating weighted random numbers corresponding to known probability distributions determined by the optical depth $\tau$ and albedo $a$ of the CSM \citep{whitney_2011}. In addition to the central star, we also consider the CSM itself as a ``distributed" source of photons. This capability is one of the strengths of the MCRT method. We describe the dust emission in Section~\ref{dust_emission}.


For each photon, \textit{SLIP} performs the numerical optical depth integration described in \citet{Code_1995}, \citet{whitney_2011}, and \citetalias{Shrestha_2018}. 
The photon's Stokes parameters are updated after each scattering event by applying the standard Mueller matrix multiplication \citep{Chandrasekhar_1960,Code_1995,whitney_2011}. After a number of scattering events depending on optical depth, a photon exits the simulation. We combine the Stokes parameters for all photons in each output bin; a single simulation from \textit{SLIP} thus yields Stokes vectors at all viewing angles ranging from $i=0-180\degr$. We normalise the summed Stokes vectors in each bin by wavelength to ensure the output fluxes have the correct units. We calculate the uncertainties in the Stokes parameters in each bin by taking the standard deviation of that parameter over all $N$ photons in the bin and normalising it to $\sqrt{N}$ (\citealt{Wood_96a,whitney_2011}; Paper I). 

\begin{figure*}
\includegraphics[width = \linewidth]{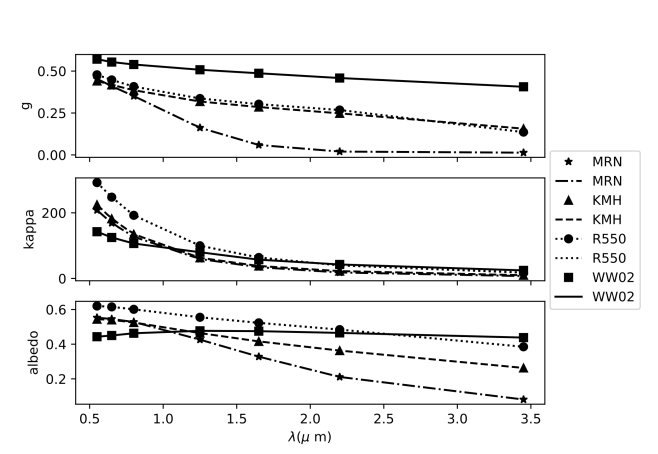}
\caption{Properties of the dust models we use in the \textit{SLIP} simulations. \textit{Top to bottom:} dust scattering asymmetry $g$, opacity $\kappa$, and albedo $a$ as functions of wavelength. Solid lines connecting the symbols represent one-dimensional interpolation of the data points.}
\label{dust_prop}
\end{figure*}
\begin{figure*}
\includegraphics[width = \linewidth]{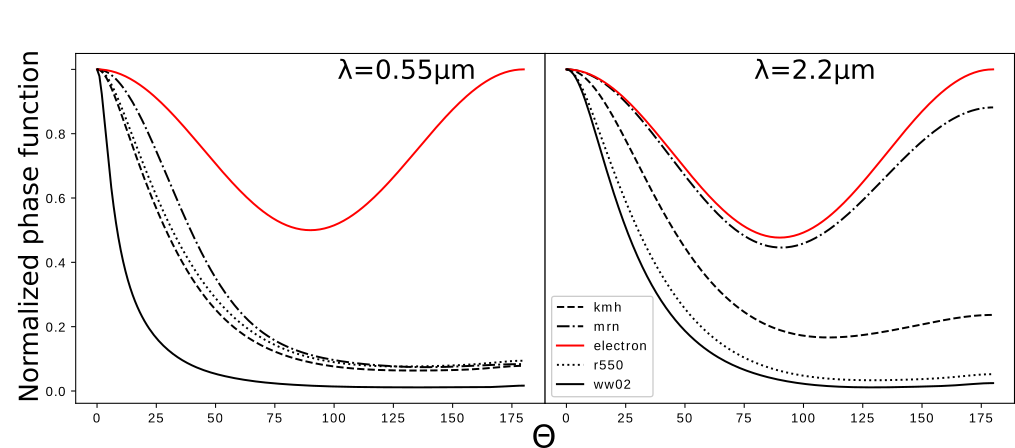}
\caption{Normalised scattering phase function for the dust models considered here (\S~\ref{method}) compared with  electron scattering (\textit{red solid lines}). The angle $\Theta$ separates the incoming and outgoing photon directions. We show two representative wavelengths corresponding to $V$ band (\textit{left}) and $K$ band (\textit{right}).}
\label{phase}
\end{figure*}
\begin{figure*}
\includegraphics[width = \linewidth]{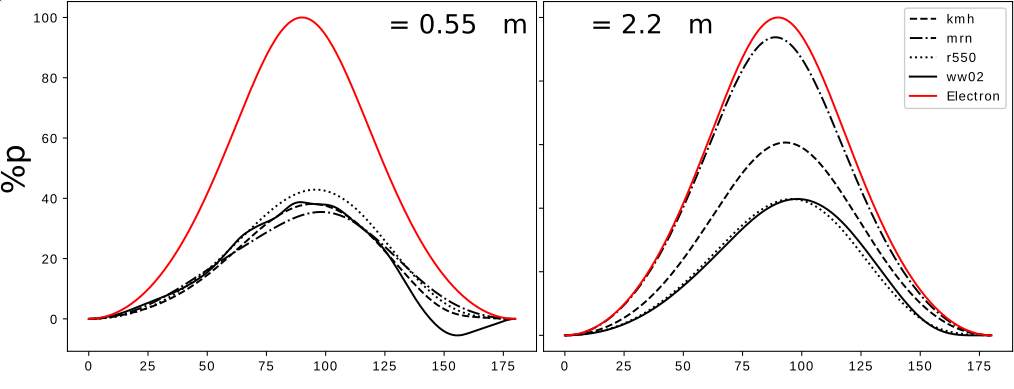}
\caption{Polarization degree as a function of scattering angle for the dust models considered here (\S~\ref{method}) compared with  electron scattering (\textit{red solid lines}). The angle $\Theta$ separates the incoming and outgoing photon directions. We show two representative wavelengths corresponding to $V$ band (\textit{left}) and $K$ band (\textit{right}). For WW02 dust, the polarization becomes negative at some large viewing angles, signifying a change in the position angle \citep{Zubko_2001}.}
\label{phase2}
\end{figure*}

Within \textit{SLIP}, we use tabular functions as described in \citet{whitney_2011} to define the scattering properties for several different dust models. The data files available from the code distribution in \citet{whitney_2013} contain the elements of the phase scattering matrix and other optical properties for several common dust models as a function of wavelength.
We list the dust models considered here along with some of their properties in Table \ref{tab:dust_types}, assuming a representative wavelength of 2.2 $\mu$m ($K$ band). In Fig.~\ref{dust_prop} we display for each dust model the variation of the dust scattering asymmetry $g$, which has larger values for more forward-throwing phase functions (forward-throwing means the probability of scattering angle being less than $90\degr$ is higher), the opacity $\kappa$, and the albedo $a$ with wavelength (corresponding to the central wavelengths of the standard Johnson filters;  \citealt{Johnson_1953}). In Figs.~\ref{phase} and \ref{phase2}, we graph the scattering probability and polarization degree as a function of scattering angle for these dust models; following \citet{Chandrasekhar_1960} and \citet{whitney_2011}, we use $\Theta$ to represent the scattering angle. We depict results for $V$ and $K$ band (0.55 and 2.2~$\micron$m, respectively) to show the range of dust behavior and to recognize that most polarization observations of bow shocks have been obtained in $K$ band. All the dust models we use have strongly forward-throwing scattering behavior at most wavelengths; however, at longer wavelengths the MRN phase function gains a significant back-scattering component and resembles that of electron scattering (Fig.~\ref{phase}). As with electron scattering, all dust models achieve the maximum polarization degree for $\sim90$\degr scattering angles, though the peak shifts to slightly larger angles for most dust types. Dust scattering also produces lower polarization overall than electron scattering. At longer wavelengths, scattering by MRN dust again closely resembles the behavior of electron scattering.

\begin{table*}
\caption{Properties of the dust models implemented within \textit{SLIP} for a representative wavelength of 2.2 $\mu$m ($K$ band). The quantity $g$ is the scattering asymmetry parameter, $\kappa$ the opacity, and $a$ the albedo.}
\label{tab:dust_types}
\centering
\begin{tabular}
{|c|c|c|c|l||l|}
 \hline
Dust type & $g$ & $\kappa$ (g/cm$^2$)  & $a$ & composition& Literature\\
 \hline
MRN & 0.02 & 18.35 & 0.211 & silicate and graphite&\citet{Mathis_1977} \\
KMH & 0.24 & 22.52 & 0.363 & silicate and graphite&\citet{Kim_1994} \\
R550 & 0.26 & 39.83 &0.483 & silicate and graphite & \citet{Clayton_2003} \\
WW02 & 0.49 & 42.70 &0.465 & silicate and amorphous carbon& \citet{Cotera_2001}\\

 \hline
\end{tabular}
\end{table*}

MRN dust is based on the standard interstellar grain model created by \cite{Mathis_1977} using a fit to observed interstellar extinction in the wavelength range $0.11-1~\mu$m. The dust particles are spherical, uncoated, and composed of graphite and silicate, with a size distribution given by $n(s) \propto s^{-3.5}$, where the grain radius $s$ ranges from $0.005~\mu$m to $0.25~\mu$m. KMH dust represents interstellar grains using a fit to observed interstellar extinction in the wavelength range of $0.1-5~\mu$m \citep{Kim_1994}. The composition and shape of the KMH dust grains is the same as in the MRN model; however, the MRN model has a sharp dust size cutoff at $0.25~\mu$m, while the KMH model dust grain size decreases smoothly with wavelength from $0.2~\mu$m to $1.0~\mu$m. For both MRN and KMH models, the ratio of total to selective extinction ($R_V$) is $3.1$. 

The WW02 dust model was obtained by extinction curve fitting to the disk of the T Tauri star HH 30 \citep{Cotera_2001}. The WW02 dust grains are larger than the MRN and KMH grains 
by a factor of approximately 2.1, and are composed of silicate and amorphous carbon. The R550 model comes from a maximum entropy method fit to HD 37022 in the Orion Nebula, with $R_V=5.5$ in the wavelength range of $0.125-3~\mu$m; it assumes a graphite and silicate composition \citep{Clayton_2003}.

We chose to use the KMH and MRN dust models because they are most representative of ISM dust properties. Hydrodynamical simulations have shown that ISM dust can be present in the bow shocks created by stellar winds of massive stars, because these dust grains can penetrate deep into the bow shock due to their inertia \citep{Marle_2015}. The WW02 and R550 models, which have larger dust grains, allow us to investigate the impact of grain size on the resulting polarization. In addition, for evolved, fast-moving red supergiants, \citet{Marle_2011} showed that larger dust grains from the stellar wind can also penetrate into the bow shock region; these larger dust grains may affect the polarization signatures of the bow shocks we study.

In this study we simulate the polarization due to dust scattering for the case of a generalised bow shock, rather than a particular object as in \cite{Neilson_2013}, \cite{Shahzamanian_2016}, and \cite{Zajacek_2017}. We describe the CSM in our models using an axisymmetric bow shock defined analytically by \cite{Wilkin_1996}. The Wilkin formulation assumes a spherically symmetric stellar wind and a locally uniform ISM, and derives expressions for the shape, mass surface density, and velocity flow in an infinitesimally thin axisymmetric bow shock. These properties of the bow shock are determined by the stellar mass loss rate, the speed of the star through the ISM, and the local ISM density.

The shape of the theoretical bow shock, which arises from 
momentum conservation and force balance considerations, is given by

 \begin{equation} 
   R(\theta)= \sqrt{3}R_0 \csc \theta \, \sqrt{1-\theta \cot\theta}
   \label{radwill}
  \end{equation}
  
\noindent \citep{Wilkin_1996}. We use this equation to define the shape of the scattering region in the \textit{SLIP} code. The standoff radius $R_0$ is defined as the location along the star's path at which the ram pressure of the ISM and stellar wind are equal; it is given by
\begin{equation} 
   R_0=\sqrt[]{\frac{\dot m_w V_w}{4\pi \rho_{I} V_\star^2}}
   \label{standoff} .
  \end{equation}

\noindent $R_0$ depends on the stellar mass-loss rate ($\dot m_w$), the density of the ISM ($\rho_I$), the stellar velocity ($V_{\star}$), and the stellar wind speed ($V_w$; \citealt{Wilkin_1996}). We initially created various dust scattering models with a range of $R_0$ values and found that the value of $R_0$ does not affect the resulting polarization when there is no scattering material interior to the bow shock. This result was also found by \citet{Neilson_2013}. In the representative models presented here, we chose $R_0=1.4$ AU, the value we used in \citetalias{Shrestha_2018}, to give a convenient scale to our simulations. Although this is a smaller radius than normally observed in stellar wind bow shocks, the fact that the polarization is insensitive to $R_0$ means we can still use our model results to derive physical parameters for real scenarios. In a study comparing \textit{SLIP} models with observations, $R_0$ could be set to a measured value or adjusted as an input parameter in order to derive physical quantities such as stellar wind speed or ISM density (\S~\ref{comparison_observation}). We also consider the case of dust within the bow shock (\S~\ref{dust_in}), in which the standoff radius affects the polarization by changing the interior optical depth.

Because of the discrete grid structure used in \textit{SLIP}, the code cannot simulate an infinitesimally thin bow shock. Thus, we assign the bow shock a radial thickness $\Delta R(\theta)$ and calculate its volume density using the thickness and mass surface density given by \cite{Wilkin_1996}. The details of this implementation can be found in \citetalias{Shrestha_2018}. We tested various thickness values 
and found no significant effects on polarization within the physically thin regime, i.e., $\Delta R(\theta) < 0.5\,R(\theta)$. Thus we choose $\Delta R(\theta) = 0.25\,R(\theta)$, which ensures the radial extent of the CSM covers at least one grid cell. The volume density is then given by 

\begin{equation}
    \rho(\theta) = \rho_I\,\frac{R_0b(\theta)}{2\Delta R(\theta)} \left\lbrace \frac{[2 \alpha (1-\cos\theta)+\tilde{\varpi}^{2}]^{2}}{\tilde{\varpi} \sqrt{(\theta-\sin\theta \cos\theta)^{2}+(\tilde{\varpi}^{2} - \sin^2\theta)^{2}}}\right\rbrace 
    \label{rho}
  \end{equation}
  
\noindent \citepalias{Shrestha_2018}. Here $\rho_I$ represents the density of the ISM, $\tilde{\varpi}^{2} = 3\,(1-\theta \cot\theta)$, and $b(\theta)$ is a geometrical function that accounts for the angular dependence of the radius, which we derived in \citetalias{Shrestha_2018}. The factor $\alpha$ is the ratio of the star's translational velocity to its wind speed; for most observed bow shocks, $0<\alpha<1$ (e.g., \citealt{Kobulnicky_2018}; \citetalias{Shrestha_2018}). Hot stars typically have low $\alpha$ values, while for cool stars, $\alpha$ may approach or even exceed 1 \citep[e.g.,][]{Ueta_2008,Mohamed_2012}. We assume $\alpha=0.1$ for an intermediate representative case and direct comparison with the simulations in \citetalias{Shrestha_2018}. When all other model parameters are fixed, larger values of $\alpha$ effectively increase the density throughout the scattering region, leading to higher overall polarization. Thus, in models of specific bow shocks, this parameter should be tuned to match the properties of the central star (\S~\ref{comparison_observation}).

Because our simulation grid has a finite size, we truncate the bow shock for large values of $\theta$. Instead of taking the density abruptly to zero, we set a cutoff angle $\theta_{\rm c}$ after which the density falls off exponentially:
 \begin{equation}
    \rho(\theta > \theta_{\rm c}) = \rho(\theta_{\rm c}) \exp[-(\theta-\theta_{\rm c})/\delta\theta_0]\; .
    \label{exp}
\end{equation}

\noindent In this expression, $\delta\theta_0$ is a constant angle governing the steepness of the density decline. After testing the effects on the model polarization of different values of $\delta\theta_0$ and $\theta_{\rm c}$, we chose $\theta_{\rm c}=2.1$ rad ($122^{\circ}$) and $\delta\theta_0=0.3$ rad ($17^{\circ}$) for all the models shown hereafter (as in \citetalias{Shrestha_2018}). 

We use the volume density $\rho(\theta)$ determined by Equations~\ref{rho} and \ref{exp} to calculate the variation of the optical depth $\tau(\theta)$ in the bow shock depending on the input optical depth $\tau_0$. Fig.~\ref{emission} (top) shows the resulting optical depth and density behavior with respect to $\theta$. The angular dependence of the density causes the optical depth to increase gradually as a function of $\theta$, then decrease steeply after the cutoff angle.

\subsection{Dust emission} \label{dust_emission}

To understand the polarization behavior in cases of dust scattering, we need to account for the effect of emission from the dust itself; we present models with dust emission in \S~\ref{sec:dust_em}. To calculate the dust luminosity within the code, we let $j_\nu$ be the dust emissivity and $dV$ a differential element of volume within the CSM. Then the dust luminosity is given by

 \begin{equation}
  L_\nu=\int j_\nu \, dV \, .
 \end{equation}

\noindent For isotropic emission, any coordinates can be used to obtain the same result. The total number of photons generated is also independent of the coordinate system. We thus choose convenient, locally normal coordinates. Let $dl$ be a thickness in the normal direction and $d\Sigma$ be a tangential area element; then we have

\begin{equation}
 L_\nu=\int j_\nu \,dl\, d\Sigma\, .
\end{equation}

\noindent The emissivity is of the form $j_\nu$ = $\kappa_\nu \rho B_\nu (T)$, where $\kappa_\nu$ is the dust opacity, $\rho$ is the mass density, and $B_\nu (T)$ is the Planck function at temperature $T$. Now we have

\begin{equation}
  L_\nu=\int \rho \,dl\, \int \kappa_\nu\, B_\nu\ (T)\, d\Sigma\, .
\end{equation}

\noindent The first integral is just the  mass surface density $\sigma$:

\begin{equation}
  L_\nu=\int \kappa_\nu \, B_\nu (T)\, \sigma\, d\Sigma\, .
\end{equation}

\noindent If we adopt the assumption that the dust grains are isothermal, in line with our LTE approximation, then the luminosity becomes

\begin{equation}
  L_\nu=\kappa_\nu \, B_\nu (T)\, \int \sigma\, d\Sigma\, .
  \label{lumi}
\end{equation}

\noindent 
With the area element given by $d\Sigma=2\pi\, R^2(\theta') \,d\theta'$, the integral 
$\int_{0}^{\theta} \sigma\, d\Sigma$
yields the total dust mass $M(\theta)$ in the bow shock up to the polar angle $\theta$.

Now Eq.~\ref{lumi} reduces to 

\begin{equation}
  L_\nu (\theta)=\kappa_\nu \, B_\nu (T)\,M(\theta)\,,
  \label{eq:dust_emission}
\end{equation}

\noindent giving the total dust luminosity up to the angle $\theta$.

\begin{figure*}
  \includegraphics[width=\textwidth]{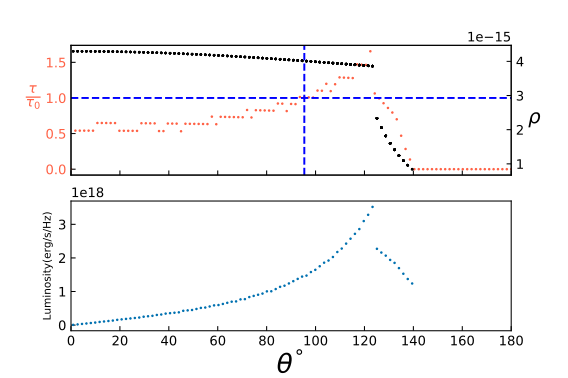}
  \caption{\textit{(top)} Variation in density (in g/cm$^3$; \textit{black points}) and normalized optical depth (\textit{red points}) within our simulated bow shock as a function of polar angle. The dashed blue lines represent the input reference optical depth $\tau_0$ at the reference angle $\theta_0$ (\S~\ref{method}). \textit{(bottom)} Dust luminosity as a function of polar angle for KMH dust with a reference temperature of 1000 K at a representative wavelength of $2.2\,\mu$m (\S~\ref{dust_emission}).}
  \label{emission}
\end{figure*}

In the lower panel of  Figure \ref{emission}, we illustrate this derived dust luminosity as a function of $\theta$ for representative parameters 
$T=(1000~\textrm{K}) \sqrt{\frac{R_0}{r}}$
(choosing $T_d=1000$ K for maximum dust emission; \S~\ref{method}), $\lambda = 2.2\,\mu$m, and $\kappa=22.52$ cm$^2$/g (corresponding to the KMH dust model; Table \ref{tab:dust_types}). 
We implement this dust emission within \textit{SLIP} by first calculating the ratio of the dust luminosity $L_\nu(\theta)$ to that of the central star $L_\nu(*)$ at each wavelength. We calculate the latter via
\begin{equation}
  L_\nu (*) = 4 \pi^2 R_{*}^2 B_\nu (T_*)~.
  \label{eq:lum_star}
\end{equation}

\noindent We then use the fraction $L_\nu (\theta)/L_\nu (*)$ to determine the number of photons emitted from the bow shock, given a certain number emitted from the star. These photons emitted from the bow shock can scatter, become absorbed, or  escape in the same way as the photons emitted from the star. In the simulations presented here, we take the central star to have the blackbody spectrum of Betelgeuse, with $R_{*} = 1000~R_\odot$ and $T_*=3500$ K \citep{Smith_2009}. We also investigated hotter central stars ($\zeta$ Oph, with  $R_{*} = 7.2~R_\odot$ and $T_*=31,000$ K, and HD 77207, with  $R_{*} = 50~R_\odot$ and $T_*=11,000$ K; \citealt{Kobulnicky_2019}). In these cases, the smaller stellar radii cause $L_\nu(*)$ to be lower than for Betelgeuse, and therefore dust emission dominates the polarization, as in the high-temperature dust cases discussed in \S~\ref{sec:dust_em}.

\section{Model Comparisons}
\label{comp}
\subsection{Comparison with electron scattering case}
\label{escatt}
We first compared the results for the intrinsic polarization produced in the bow shock by several dust models with that produced by electron scattering, with photons emitted from the central source only in both cases. For the electron-scattering models, we chose a representative CSM temperature of 10,000 K; we found in \citetalias{Shrestha_2018} that the temperature of the bow shock does not significantly affect the results of electron-scattering simulations when albedo is constant and temperature is greater than 10,000 K. In the electron-scattering models, we fixed the albedo at 0.544 and 0.464 to compare with dust at 0.55 $\mu$m and 2.2 $\mu$m, respectively. We chose these albedo values to match those of KMH dust, but the other dust types we consider have similar albedo values at these wavelengths (Fig.~\ref{dust_prop}).

\begin{figure*}
\includegraphics[width = \linewidth]{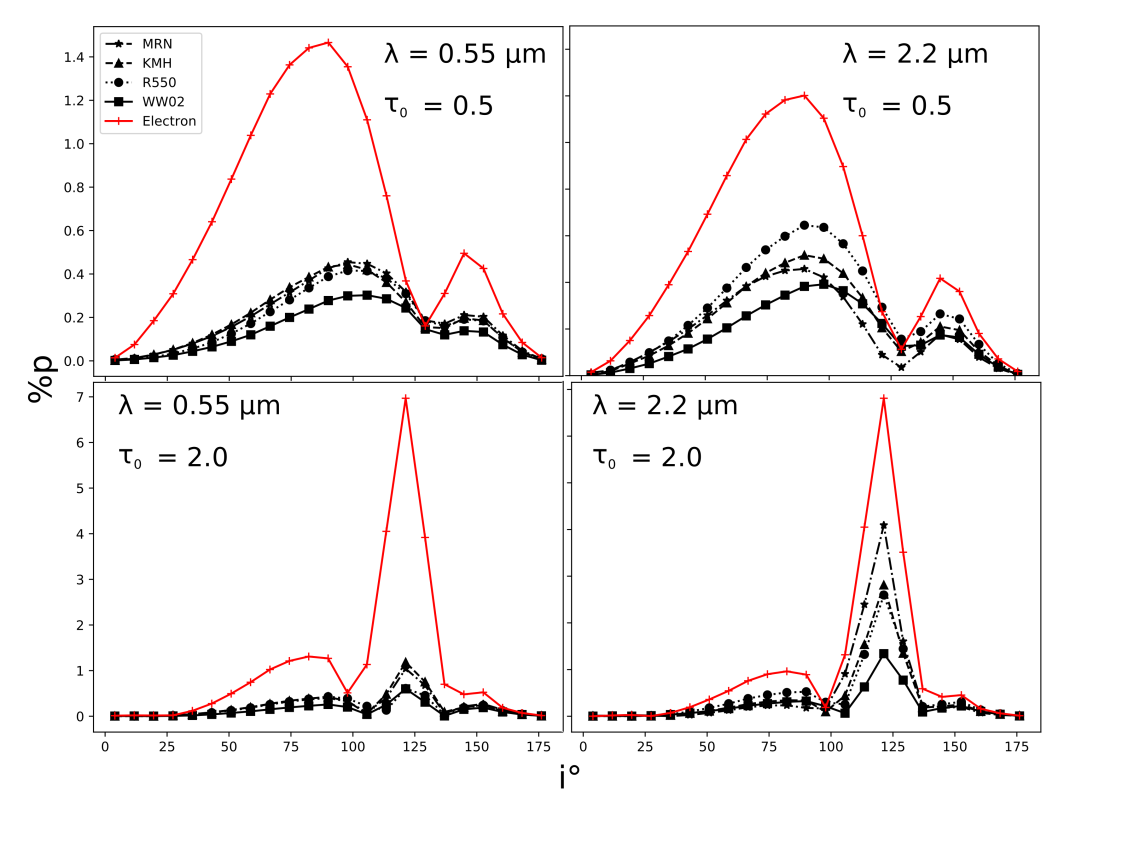}
\caption{Simulated polarization as a function of inclination angle for different dust types compared with electron scattering (\textit{red plus signs}). We display two different optical depths and two different wavelengths. Albedo for all cases is 0.544 for the $0.55\,\mu$m ($V$ band) models and 0.464 for the $2.2\,\mu$m ($K$ band) models (\S~\ref{escatt}). Error bars are smaller than the plotted points.}
\label{electroncomp}
\end{figure*}

Fig.~\ref{electroncomp} compares the variation in polarization with viewing angle between electron scattering and several dust types for two different optical depths and two representative wavelengths ($\lambda=0.5$ and 2.2\,\micron\, corresponding to $V$ and $K$ bands). In all cases, the amount of polarization is higher for electron scattering than for dust scattering; this can be explained by referring to the scattering behavior of these mechanisms depicted in Figs.~\ref{phase} and \ref{phase2}. The electron case has the highest probability of scattering at $\Theta=90\degr$, which polarizes the light by $100 \%$. The dust models have both a lower probability of 90\degr~scattering and a lower maximum polarization degree (though MRN dust approximates the electron-scattering behavior in $K$ band, as noted in \S~\ref{method}). 

Regardless of wavelength, all dust models show a qualitatively similar behaviour to electron scattering, although with a first polarization peak that is shifted to larger angles than 90\degr. This behaviour is a consequence of the dust polarization functions shown in Fig.~\ref{phase2}. The dip in polarization near 130\degr~at lower optical depths (top panels) occurs because the albedo for these models is $<1$, so absorption effects are important. At larger viewing angles, our line of sight goes along the arms of the bow shock, corresponding to a longer path length and greater possibility of absorption. At the higher optical depth (bottom panels) we see a polarization peak corresponding to this same angle. As with the central-source electron-scattering case \citepalias{Shrestha_2018}, the increasing rate of multiple scattering causes increased contributions from negative Stokes $q$ values. This flip in $q$ from positive to negative also produces a rotation in position angle $\Psi$ from 0\degr~to 90\degr~at these viewing angles. 

Among our dust models, the MRN model has a phase function and polarization behaviour most similar to electron scattering, particularly at $\lambda=2.2\,\mu$m (Figs.~\ref{phase},  \ref{phase2}). The WW02 model is the most different from the electron scattering case and also from the other dust models, since it is extremely forward-throwing.  Fig.~\ref{electroncomp} shows that these differences in scattering and polarizing behaviours translate into clear differences in model results, as we will explore further in the next sections.

\subsection{Dust parameter study}
\label{dustpar}

\begin{figure}
  \includegraphics[width=\columnwidth]{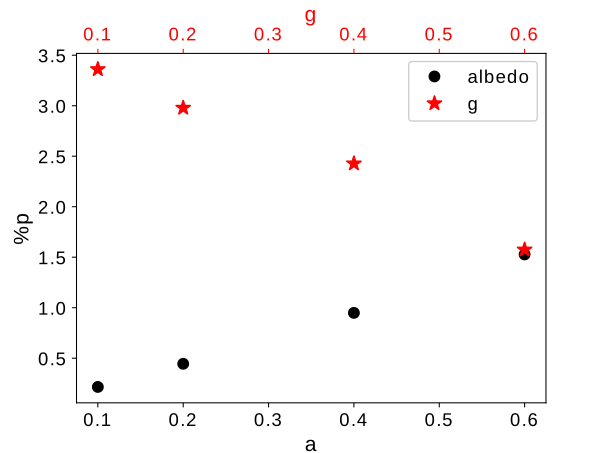}
  \caption{Variation in $K$-band polarization as a function of albedo $a$ and dust asymmetry factor $g$ in the Henyey-Greenstein function (\S~\ref{dustpar}), with all other model parameters held constant in each case. We show results for a representative viewing angle of $i=90\degr$. Error bars are smaller than the plotted points.}
  \label{pva-g}
\end{figure}

To investigate the separate effects of the dust albedo $a$ and asymmetry factor $g$ on the polarization, we created several test simulations using a range of each of these values in the general Henyey-Greenstein dust scattering function \citep{Henyey_1941}: 
\begin{equation}
    p(\Theta) = \frac{1}{4 \pi}\frac{1-g^2}{(1+g^2-2g \cos{\Theta})^{3/2}}
    \label{hgfunc}
\end{equation}

\noindent For these models we used the following parameter values in the scattering matrix \citep{whitney_2011}: we set the dust opacity $\kappa$ to $22.52$ cm$^2$/g (the value for KMH dust at $K$ band), the maximum linear polarization $p_l$ to 1, the maximum circular polarization $p_c$ to 0, and the skew factor $s$ to 1. For all test models, we set the optical depth to $\tau_0=0.5$ and did not allow the dust to emit.
When varying $g$, we fixed $a=1$, and when varying $a$, we fixed $g=0.44$. 

The resulting polarization increases linearly with the dust albedo $a$ and decreases linearly with the dust asymmetry factor $g$ (Fig.~\ref{pva-g}). These results will help us interpret the polarization results we present below. 

\section{Model Predictions from \textit{SLIP}}\label{results}
We describe the geometrical and density setup of our simulations in Section~\ref{method} above; further details are given in \citetalias{Shrestha_2018}. For the models presented here, unless otherwise specified, we used $R_0 = 1.4$ AU and $\alpha=0.1$ to create a representative stellar wind bow shock comparable with previous simulations (we discuss the physical implications of these choices in \S~\ref{method}). We also set the maximum extent of the grid at $R_{\textrm{max}} = 6.68$ AU, the cutoff angle at $\theta_c=2.1$ rad ($122^{\circ}$), and the exponential decay factor at $\delta\theta_0=0.3$ rad ($17^{\circ}$).  All these parameters have the same values as in \citetalias{Shrestha_2018}.

We used the University of Denver's high-performance computing cluster (HPC) to create most of the simulations; this cluster consists of Intel Xeon processors with 456 computational cores running at 2.44 GHz. Each of our model runs used 16 CPUs with $10^8$ photons per CPU. With 23 output bins in polar angle, this yielded polarization uncertainties on the order of $\sigma_p(\%) \sim 0.01$ per bin. Each run took $\sim 60-70$ minutes for completion, with slightly longer times for larger values of $\tau_0$. We also used the Stampede supercomputer at the Texas Advanced Computing Center (TACC) for some of the higher optical depth simulations.

We created simulations both with and without emission from the dust in the bow shock. For both these cases, we studied the impact of the dust size and composition on the polarization behaviour for resolved and unresolved bow shocks. To simulate resolved bow shocks, we retain the spatial information contained in the outgoing photons. To simulate unresolved bow shocks, we combine all outgoing photons in a particular polar angle bin to calculate a single set of polarization values. We also studied the impact of varying the dust temperature and optical depth. For the simulations without dust emission, changing the dust temperature does not change the polarization behaviour because the opacity and albedo are fixed for each dust model at each wavelength (Table~\ref{tab:dust_types}; Fig.~\ref{dust_prop}). We present the results for these different scenarios below. We studied optical depths from $\tau_0=0.5-2.0$ in detail. At higher optical depths, our tests show that the overall polarization level typically increases beyond what we display here, but with much lower polarized intensity due to significant absorption. In  these simulations, we found polarization position angles very close to $\Psi=0\degr$ for unresolved cases (in the model coordinate system of \citetalias{Shrestha_2018}; this direction is tangential to the shock at the bow head) for most parameter combinations and viewing angles. Similarly, resolved cases produce the centrosymmetric position angle pattern expected when polarization peaks near $90\degr$ scattering angles. Thus, for simplicity we display the results in per cent $p$ only. 

We also note that these models do not simulate the polarization created by light scattering in the intervening dust between the observer and the bow shock, as discussed in Section \ref{sec:intro}, because we are concerned with polarization intrinsic to the stellar system. Galactic dust properties vary with location on the sky, and thus interstellar polarization (ISP) effects are more appropriately addressed via observations of specific objects of interest. In the sections below, we discuss interpretation of our model results with and without ISP estimates.

\subsection{Models without dust emission}

We first present results without dust emission, for four dust types at four representative inclination angles, for both resolved and unresolved bow shocks. 

\subsubsection{Dust type dependence -- resolved bow shock}
\label{result_noemission_r_opt}

\begin{figure*}
\includegraphics[width=\textwidth]{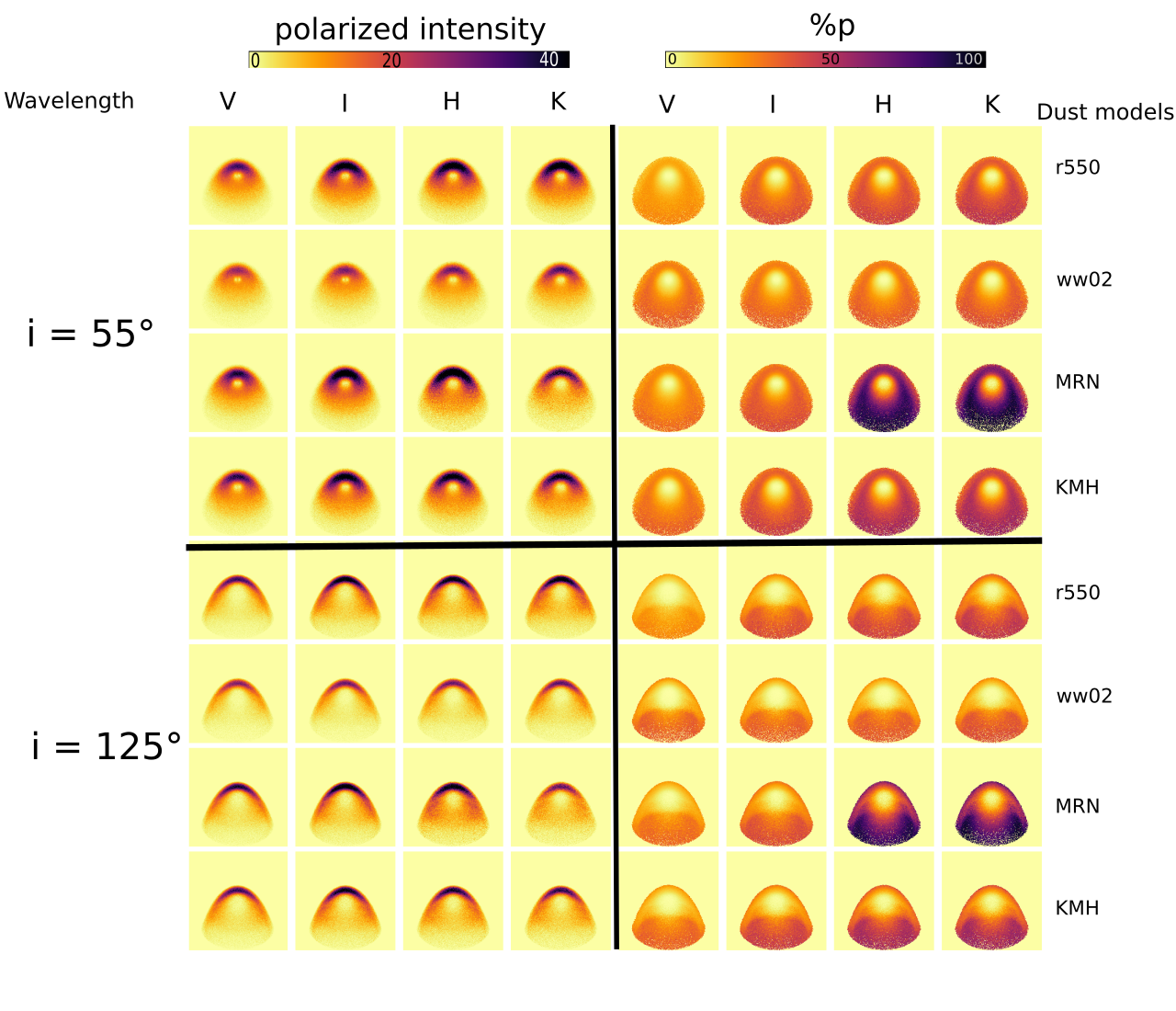}
\caption{Simulated maps of polarized intensity (\textit{left}) and polarization (\textit{right}) for a \textit{SLIP} model with no dust emission and a reference optical depth of $\tau_0=0.5$ (Fig.~\ref{result_noemission_r_opt}). Intensities are in arbitrary units. We describe the different dust types in \S~\ref{method}.}
\label{map_othin}
\end{figure*}

In Fig.~\ref{map_othin}, we present maps of polarized intensity and fractional polarization (expressed as a percentage) for resolved bow shocks with $\tau_0 = 0.5$. We show results for four different dust types in four wavebands 
at two representative inclination angles symmetric around the $z=0$ plane, $i=55\degr$ and $i=125\degr$. We calculate polarized intensity by multiplying fractional $p$ by intensity; this quantity represents the polarized light arising from the system.

In all cases shown in  Fig.~\ref{map_othin}, the polarized intensity is concentrated near the bow head, where the dust density is highest. This behaviour is similar to that of the electron-scattering simulations in \citetalias{Shrestha_2018} in the low optical-depth regime, and reflects the peak in polarization degree near $90\degr$ shown in Fig.~\ref{phase2}. For the R550 and WW02 dust models (which have larger grains), the absolute polarized intensity increases with wavelength for the shorter wavelengths and remains constant at $H$ and $K$. However, for KMH and MRN dust (smaller grains), the polarized intensity increases only up to $H$ band and decreases at $K$. This behaviour is due to the low albedos of MRN and KMH dust in the $K$ band (Fig.~\ref{dust_prop}; Fig.~\ref{pva-g}). For R550 and WW02 dust, the albedo is almost constant, but the \textit{g} parameter decreases slightly with wavelength, causing almost constant polarized intensity.

We also see from Fig.~\ref{map_othin} that the overall fractional polarization increases with increasing wavelength for all dust models at both inclination angles. This is similar to the behaviour of the polarized intensity; however, the fractional polarization does not decrease at longer wavelengths for any of the dust models we used. This is because the general decrease in albedo with wavelength causes a decrease in the the total intensity (polarized and unpolarized) as more photons become absorbed. The net effect is an increase in fractional polarization with wavelength. 

For most dust models, although the polarized intensity is high at the bow head, the largest fractional polarization arises from the lower part of the bow shock (higher values of $\theta$). This is consistent with our results for the central-source electron scattering case \citepalias{Shrestha_2018}, and is a consequence of the bow shock geometry. At higher $\theta$ values, the density is low so few photons scatter, but those that do tend to become highly polarized. MRN and KMH dust create more polarization than the other dust types at long wavelengths, due to their higher peak polarization values (Fig.~\ref{phase2}).

 \begin{figure*}
 \includegraphics[width=\textwidth]{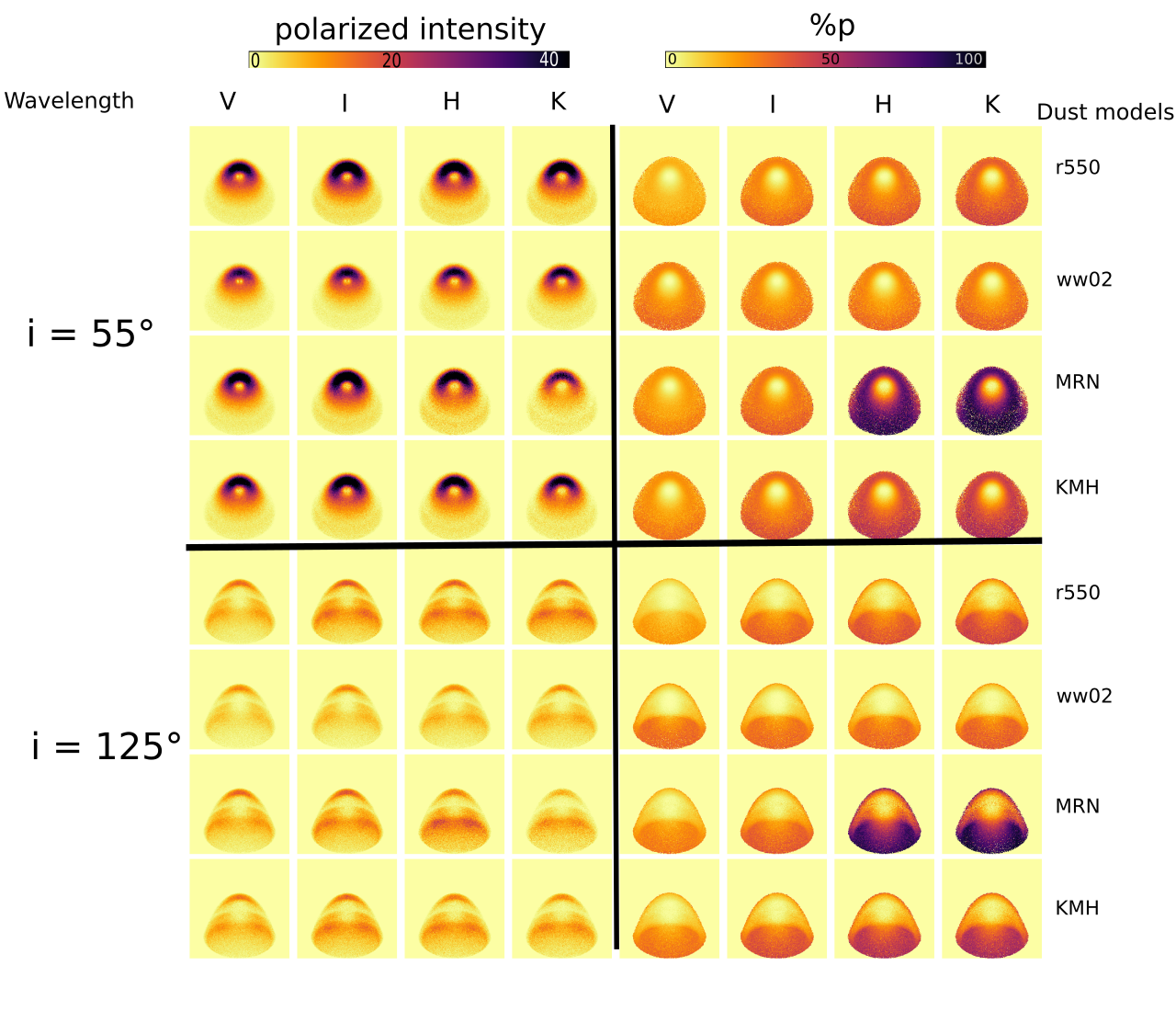}
 \caption{As in Fig.~\ref{map_othin}, but for a reference optical depth of $\tau_0=2.0$.}
\label{map_othick}
\end{figure*}

Fig.~\ref{map_othick} shows polarized intensity and polarization maps for resolved bow shocks with a higher reference optical depth of $\tau_0 = 2.0$. Unlike in the case of pure electron scattering \citepalias{Shrestha_2018}, we do not see a significant decrease in polarization at  larger optical depths. Because of the forward-throwing nature of most of these dust models (Fig.~\ref{phase}), multiple scattering does not randomize polarization vectors to the extent it does in the case of electron scattering.

The most prominent difference between the low and high optical-depth cases is the behaviour of the polarized intensity at higher inclination angles. In the case of higher optical depth and higher viewing angle, most of the polarized intensity arises from the middle portion of the bow shock, which at this viewing angle corresponds to the interior wall of the shock cone. Because these dust models are strongly forward-throwing, at higher optical depth, the photons reaching us have likely scattered twice at angles near $90\degr$ rather than backscattered once. The first scatter occurs in the high-density region in the far side of the cone near the bow head, and the second scatter redirects the photon into our line of sight. Both scatters produce polarization in the same direction, so the resulting photon is highly polarized. Since the density in the near side of the cone is lower at greater $\theta$ values, the polarized photons can easily escape from these lower density region. This viewing angle corresponds to the second peak in Fig.~\ref{electroncomp}.

At higher viewing angles, we see a ``belt" of low polarization crossing the shock cone in the polarized intensity maps. We first saw this effect in the case of electron scattering with absorption \citepalias[][Fig. 11]{Shrestha_2018}. It arises when multiply scattered photons become absorbed in the regions of highest optical depth.

The results shown in Figs.~\ref{map_othin} and \ref{map_othick} suggest that a single resolved observation of a dusty bow shock could distinguish lower from higher optical depths, but only in the case of a large viewing angle and only with a polarized intensity map; at lower angles the maps are similar in both polarization and polarized intensity. Some constraints on dust type are possible with polarization and polarized intensity images at multiple wavelengths (\S~\ref{comparison_observation}). In the case of a single polarization or polarized intensity image at a single wavelength, it is reasonable to assume that ISP will affect all parts of the image equally, so that features within the image are still reliable diagnostics.

\subsubsection{Dust type dependence -- unresolved bow shock}
\label{result_noemission_ur_opt}
\begin{figure*}
\includegraphics[width = \linewidth]{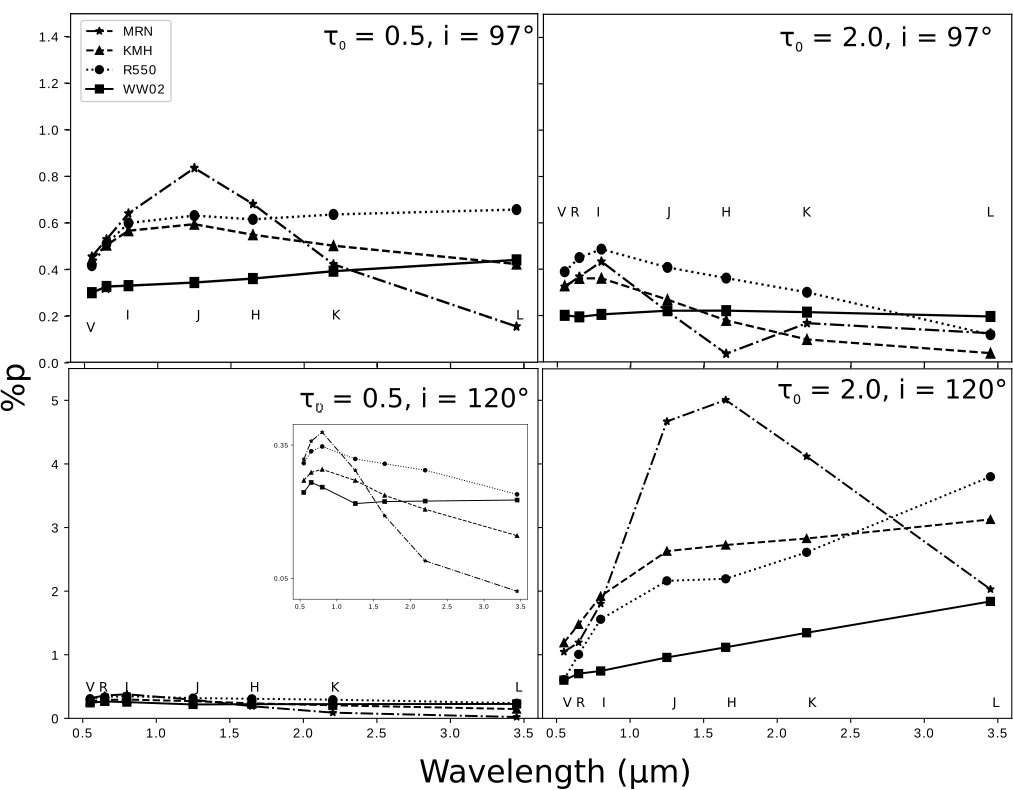}
\caption{Simulated polarization as a function of wavelength for unresolved cases (\S~\ref{result_noemission_ur_opt})) at two different inclination angles for four different dust types (\S~\ref{method}) and reference optical depths of $\tau_0 = 0.5$ and $\tau_0 = 2.0$. The dust in these models does not emit. Error bars are smaller than the plotted points. The lower left panel includes an inset showing a zoomed-in view of the polarization curves. The $x$-axis of the inset is the same as in all the other panels.}
\label{sed_diffdust}
\end{figure*}

In Fig.~\ref{sed_diffdust}, we present simulated polarized spectral energy distributions (pSEDs)
for different dust models and inclination angles in the case of an unresolved bow shock. We selected inclination angles of $i = 97\degr$ and $i = 120\degr$ because these are the two angles that produce polarization peaks for the optically thin and optically thick cases, as seen in Fig.~\ref{electroncomp}.

For the optically thin case at $i = 97\degr$ (top left panel of Fig.~\ref{sed_diffdust}), the magnitude of polarization increases with wavelength for the R550 and WW02 dust models. However, for the MRN and KMH dust models, the polarization increases up to $1.25$ $\mu$m and decreases for longer wavelengths. This behavior is reminiscent of the well known Serkowski law for  interstellar polarization, which was derived empirically from observations of Milky Way stars \citep{Serkowski_1975}. A similar trend holds for all other inclination angles. This trend can be explained by the behavior of $g$ and albedo with respect to wavelength for MRN and KMH dust types (Fig.~\ref{dust_prop}). For both these dust models, $g$ and $a$ decrease with increasing wavelength, but the albedo shows an inflection point at $1.25$ $\mu$m, decreasing more steeply for longer wavelengths. Thus, a wavelength of $1.25$ $\mu$m ($J$ band) represents an optimum point where $g$ is low, indicating that most of the scattered photons become polarized, and $a$ is still high enough to produce significant scattering.
This combination results in relatively high magnitudes of polarization. The MRN dust model has the steepest drop in $g$ between 0.55 and 1.25 $\mu$m (Fig.~\ref{dust_prop}),
and thus the largest polarization peak in its pSED (Fig.~\ref{sed_diffdust}). Similar trends hold at $120\degr$ (lower left panel of Fig.~\ref{sed_diffdust}), with the overall polarization magnitude suppressed due to the density cutoff. At this angle, because of the lower densities in the line of sight, more photons escape without scattering than at smaller angles. 

WW02 dust produces the smallest amount of polarization for all the inclination angles at wavelengths less than $1.25$ $\mu$m. This can be attributed to low values of $a$ (less scattering) combined with high values of $g$ (less polarization per scattered photon) in the short-wavelength regime compared to other dust types. We also note that the pSEDs at this angle resemble the albedo curves in Fig.~\ref{dust_prop}, except for MRN dust, which has very different scattering properties than the other dust types at longer wavelengths (Fig.~\ref{phase}).

The right-hand panels of Fig.~\ref{sed_diffdust} display pSEDs for all our dust types for $\tau_0 = 2.0$. In this case multiple scattering plays an important role in dictating the polarization behaviour; it gives rise to a second polarization peak near $120\degr$ (Fig.~\ref{electroncomp}).

In this regime, at the $97\degr$ viewing angle (upper right panel of Fig.~\ref{sed_diffdust}), the behaviour of the albedo as a function of wavelength again dominates the polarization, because photons interact with dust at higher rates and thus the value of $a$ strongly controls whether scattered photons can escape. At $120\degr$ (lower right panel of Fig.~\ref{sed_diffdust}), the line-of-sight density is low due to the exponential falloff in our bow shock model (\S~\ref{method}), so this reduces the effectiveness of the albedo. Hence at high optical depths and high viewing angles, the $g$ value affects the polarization more strongly than the albedo, and this causes an increase in polarization with wavelength for most dust types as $90\degr$ scattering becomes more common. However, for MRN dust, the albedo is so low at longer wavelengths that almost no photons escape even along this low-density line of sight.

The pSEDs in Fig.~\ref{sed_diffdust} suggest new observational techniques for constraining the properties of unresolved bow shocks. Multiple polarization observations in different wavelength bands can distinguish among the potential dust types causing the polarization. In particular, regardless of optical depth or viewing angle, the behaviour of the polarization in the $IJH$ and $KL$ regions is markedly different among dust types, so these spectral regions can serve as dust diagnostics. 

 We caution that ISP contributions may modify the observed pSED shapes. At the near-IR wavelengths we consider here, the ISP typically obeys a power law with index $\beta\approx1.6$ \citep{martin1990,martin1992}. As noted by \citet{buchholz2013}, a polarization ratio between wavebands that departs significantly from the power-law behavior is a signature of intrinsic polarization, as is a position angle rotation with wavelength. An assumed ISP function (or one constrained by observation) could easily be vectorially combined with our model results to match a given pSED.  Even a single observation in one wavelength band can provide some constraints on the type of dust producing the polarization, if the ISP is reliably estimated. Once the dust type is known or constrained, the specific value of polarization can be used to place limits on the optical depth and inclination angle. 

\subsection{Models with dust emission}
\label{sec:dust_em}

As discussed in Section~\ref{dust_emission}, we also constructed models incorporating photons emitted by the dust in the bow shock in addition to photons emitted by the star. This emission is dependent on the dust opacity and temperature as well as the photon wavelength (Eq.~\ref{eq:dust_emission}). We kept the total number of model photons constant, but allowed the ratio of photons arising from the star and the bow shock, $L_\nu/L_\nu(*)$, to change with the dust temperature.

First we investigated how varying the input dust temperature changes the polarization for different dust types at various wavelengths. Then we checked how changing the dust type changes the polarization results in cases incorporating dust emission. 

\subsubsection{Temperature dependence -- resolved bow shock}
\label{result_emission_r_temp}

\begin{figure*}
\includegraphics[width = \linewidth]{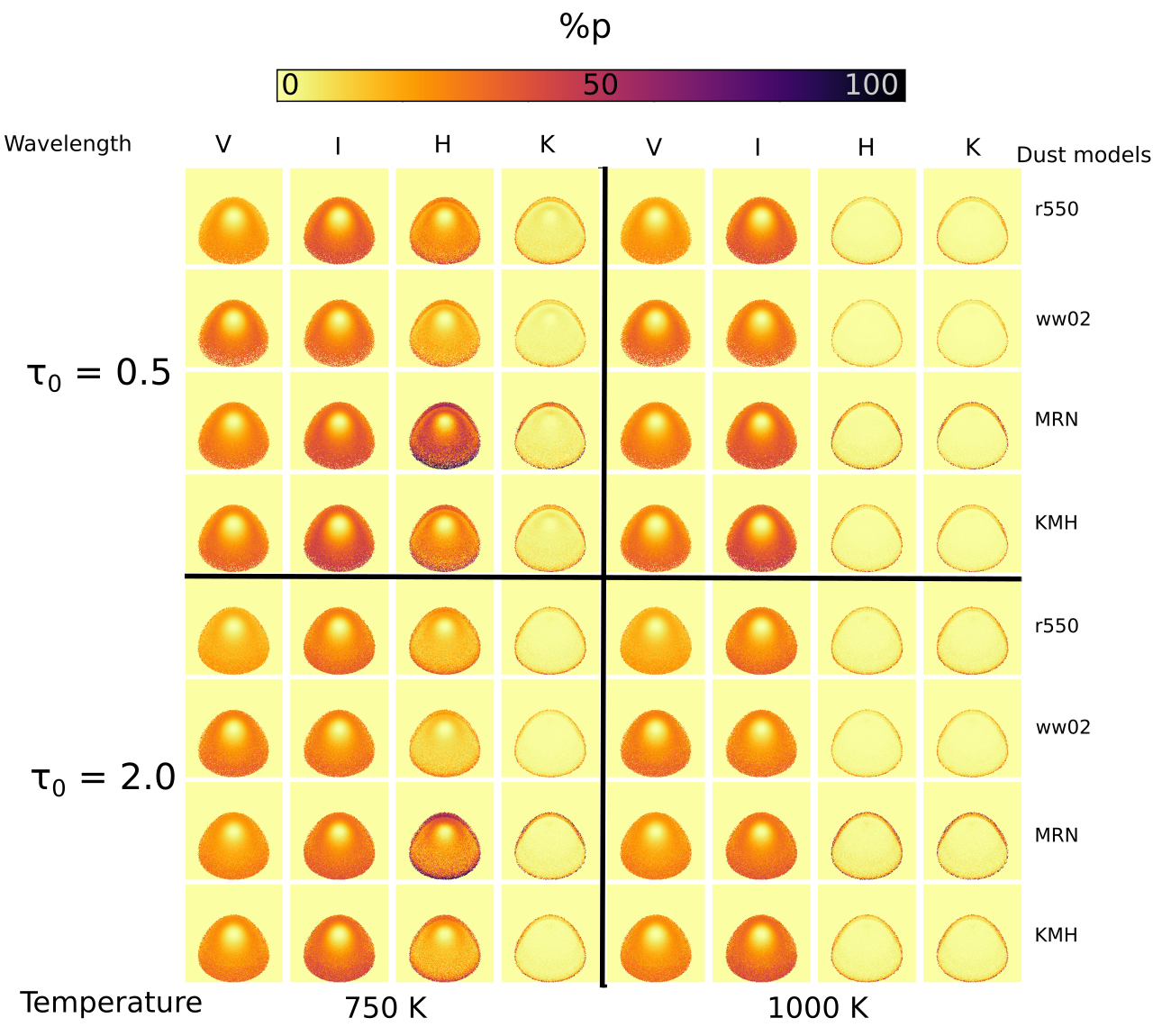}
\caption{Simulated polarization maps for a resolved model bow shock with emitting dust at $T_d$ = 750 K (\textit{left}) and 1000 K (\textit{right}). All these results are for $i = 55\degr$.}
\label{pmap_emission}
\end{figure*}

Figure~\ref{pmap_emission} shows the polarization maps resulting from our simulations of resolved bow shocks for four different dust types at four different wavelengths. We display results for two different optical depths and two different temperatures, but because we found very little variation of the polarized intensity with temperature, we present only fractional polarization maps here. We show only one representative viewing angle, $i=55\degr$. The results for $i=125\degr$ are qualitatively similar. The main differences between the two angles are the pattern of polarization and a slight overall increase in polarization at $I$ band in most cases, compared with the $55\degr$ results. These differences are also seen in Figs.~\ref{map_othin} and \ref{map_othick}.

At shorter wavelengths ($V$ and $I$ bands), the polarization maps do not vary significantly with temperature. However, at $H$ and also at $K$ for low optical depth, the polarization map is distinct between 750 K and 1000 K. (This behaviour also holds at $J$ band, which we do not display here.) At the lower temperature, there is substantially more polarization across the face of the bow shock, which disappears at 1000 K. At higher temperatures, the proportion of model photons arising from the bow shock increases. Because the Planck curve for the relatively cool dust peaks at longer wavelengths than $K$, the number of $JHK$ photons increases more than the number of $VRI$ photons. These $JHK$ photons arising from the bow shock create an effective ``distributed" photon source, whose signatures we modeled in \citetalias{Shrestha_2018}.
When photons arise from a distributed source, their polarization vectors cancel throughout most of the bow shock shape, leaving net polarization primarily at the limb and the wings far from the bow head. Thus, it should be possible to constrain the dust temperature in a resolved bow shock around a star with relatively high IR luminosity by obtaining a single polarization map in the $J$ or $H$ band, regardless of dust type or optical depth. For lower-luminosity stars, the dust emission dominates the polarization at all IR wavebands, regardless of dust temperature.

\subsubsection{Temperature dependence -- unresolved bow shock}

\label{result_emission_ur_temp}
We created simulations for different dust types at various dust temperatures, in both optically thin ($\tau_0=0.5$) and optically thick ($\tau_0=2.0$) cases. When the bow shock is optically thin, there is little polarization variation with temperature for any dust type. This is because at low optical depths, multiple scattering is not the dominant factor dictating the polarization: photons arising from the central star and from the bow shock will, on average, scatter equal numbers of times. Thus, this insensitivity to temperature agrees well with the results of \citetalias{Shrestha_2018}, where for unresolved bow shocks, we found that the amount of polarization and its behaviour as a function of viewing angle is similar for both central and distributed sources when $\tau_0 = 0.5$. 

\begin{figure*}
\includegraphics[width = \linewidth]{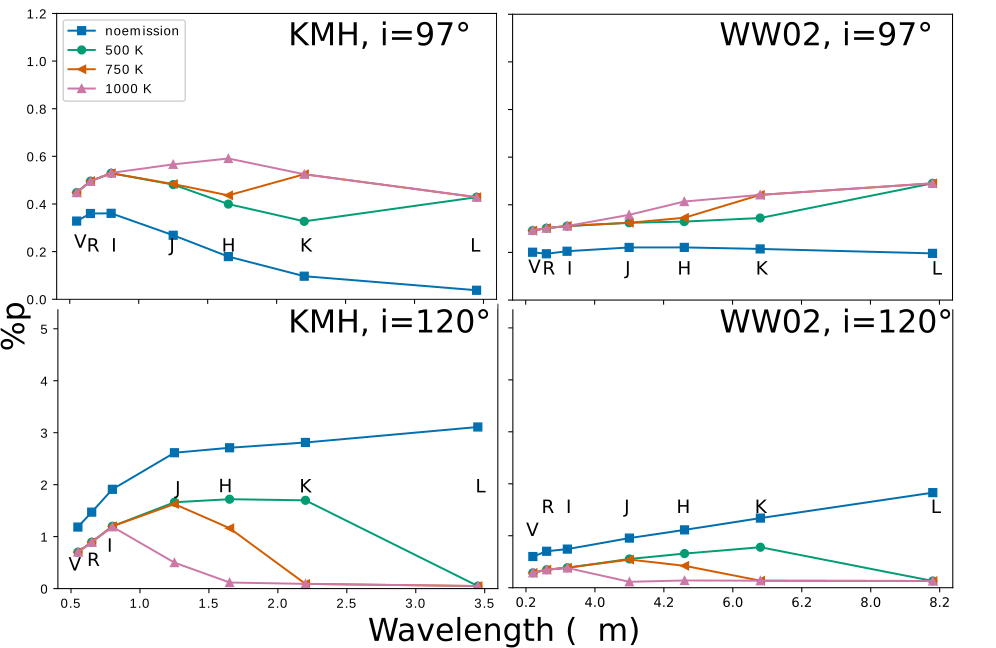}
\caption{Simulated polarization as a function of wavelength for emitting KMH (\textit{left}) and WW02 dust (\textit{right}) at two viewing angles and a reference optical depth of $\tau_0 = 2.0$. The colors and symbols represent different dust temperatures. In the ``no-emission" case, the dust in the bow shock does not emit, so all photons arise from the central star. Error bars are smaller than the plotted points.}
\label{sed_difftemp}
\end{figure*}

At high optical depths ($\tau_0=2.0$), the behaviour and amount of polarization changes with temperature for $T>500$ K. At lower temperatures than this, the dust does not emit enough photons to change the polarization. Figure~\ref{sed_difftemp} displays the variation of polarization with wavelength in the high optical-depth case for KMH and WW02 dust (as well as the case with no dust emission), at viewing angles of $97\degr$ and $120\degr$. We chose these two angles because they correspond to the peaks in polarization seen in Fig.~\ref{electroncomp}. As we found above (Figs.~\ref{phase2} and \ref{electroncomp}), polarization by dust scattering for these dust types behaves similarly to polarization by electron scattering as a function of viewing angle, but reaches a maximum at slightly greater than $90\degr$.

At $i = 97\degr$ (top panels of Fig.~\ref{sed_difftemp}), the addition of dust emission causes an increase in the degree of polarization over the case with no emission, and this increase is larger for higher-temperature dust. We attribute this behaviour to a strong increase in the fraction of photons emitted from the bow shock as temperatures rise, as discussed above. This means that the contribution of photons arising within the farther wall of the bow shock becomes increasingly important at higher temperatures. These photons must scatter in both walls of the shock to reach the viewer. The total escaping flux decreases because these photons have more chances to become absorbed, but the total polarized flux increases because the photons that do not become absorbed must scatter multiple times within the shock before escaping. The net effect is an increase in fractional polarization with temperature.

At $i = 120\degr$, multiple scattering effects create a ``secondary" polarization peak (Fig.~\ref{electroncomp}), as we also found in \citetalias{Shrestha_2018}; this angle is insensitive to the type of scattering. At this viewing angle, escaping photons scatter fewer times because of the density falloff in the arms of the bow shock. Thus, the polarized flux \textit{decreases} with temperature at this angle (for longer wavelengths) because more photons can escape the shock region without scattering. This leads to a decrease in polarization with temperature, in contrast to the $i=97\degr$ viewing angle. We found similar effects at both these viewing angles for the case of electron scattering with $a<1, \tau_0=2.0$, when we compared the ``central" and ``distributed" photon sources in \citetalias{Shrestha_2018}. 

The behaviour of the polarization with wavelength is complex, as shown by the pSEDs in Fig.~\ref{sed_difftemp}. The differences in pSED morphology as a function of viewing angle are consistent with our discussion in \S~\ref{result_noemission_ur_opt}; we see again that at the $97\degr$ viewing angle, the albedo behaviour dictates the pSED shape, while at $120\degr$, the dust asymmetry parameter is the more dominant contributor. In addition, as noted above in this section, low temperatures and high temperatures create different pSED morphologies. These differences result from the properties of the Planck curve: 1000 K dust emits most strongly in the $J$ band, while 500 K dust only contributes significant emission at $L$. When dust emission becomes important, the polarization behaviour changes because proportionally more photons arise from a spatially distributed emitting area (\S~\ref{result_emission_r_temp}). These effects are reflected in Fig.~\ref{sed_difftemp} in that for both viewing angles, all polarization curves with dust emission eventually approach the 1000-K curve as wavelength increases.

These results provide further considerations regarding the use of multi-wavelength polarization observations to constrain dust properties. Polarization magnitudes differ between the emission and no-emission cases at both angles, and the difference widens with increasing wavelength. If ISP can be estimated, even a single-wavelength observation could thus serve as a diagnostic of emitting dust in the bow shock and place limits on the dust type. However, more reliable conclusions could be drawn from observational pSEDs. Dust emission affects pSED morphology, but only at wavelengths longer than $I$ band, so $VRI$ observations may still be used to distinguish dust type (Fig.~\ref{sed_diffdust}). The longer-wavelength $IJH$ and $KL$ slopes discussed in \S~\ref{result_noemission_ur_opt} are modified by dust emission, but comparing their morphologies with those produced by the no-emission case can provide diagnostics of dust temperature and viewing angle. 
Finally, pSED morphologies will be affected by the ISP contribution, which can be be taken into account as discussed above (\S~\ref{result_noemission_ur_opt}). We explore observational dust type diagnostics in \S~\ref{comparison_observation}.

\subsubsection{Dust type dependence -- resolved bow shock}
\label{dusttype_r_emission}

As shown in Fig.~\ref{pmap_emission}, there is little difference in the polarization behavior among the four dust types at higher temperatures. However, at a dust temperature of 750 K we see variation in the polarization maps with dust type, particularly in the $H$ band. Here, MRN dust produces noticeably more polarization than the other dust types, primarily due to its rapidly decreasing $g$ value at this wavelength (Fig.~\ref{dust_prop}; Fig.~\ref{pva-g}). Additionally, for a given dust type and temperature, as the wavelength increases, the polarization concentrates near the edges of the bow shock. This occurs because longer wavelengths approach the Planck function peak at these temperatures, and thus correspond to a higher dust luminosity compared with that of the star. Overall, polarization maps seem to have limited power to diagnose the type of dust in a resolved bow shock; however, $H$- or $K$-band images could potentially distinguish MRN or KMH (ISM-type) dust from the dust types with larger grains (\S~\ref{method}).

\subsubsection{Dust type dependence -- unresolved bow shock}
\label{dusttype_ur_emission}

To investigate the polarization behaviour of different dust types in cases of unresolved bow shocks with dust emission, we created pSEDs for our four different dust types at various temperatures. At 500 K, and at higher temperatures for the lower optical depth of $\tau_0 = 0.5$, the pSEDs are very similar to those produced in the case of no dust emission (\S~\ref{result_noemission_ur_opt}; Fig.~\ref{sed_diffdust}; because of these similarities, we do not display these cases here). This result agrees well with the findings of \citetalias{Shrestha_2018}, in which we saw that for an unresolved bow shock, the amount and behaviour of polarization at $\tau_0 = 0.5$ is similar for central and distributed sources. Because the lower optical depths reproduce earlier results, we focus in this section on models with $\tau_0 = 2.0$. 

\begin{figure*}
\includegraphics[width = \linewidth]{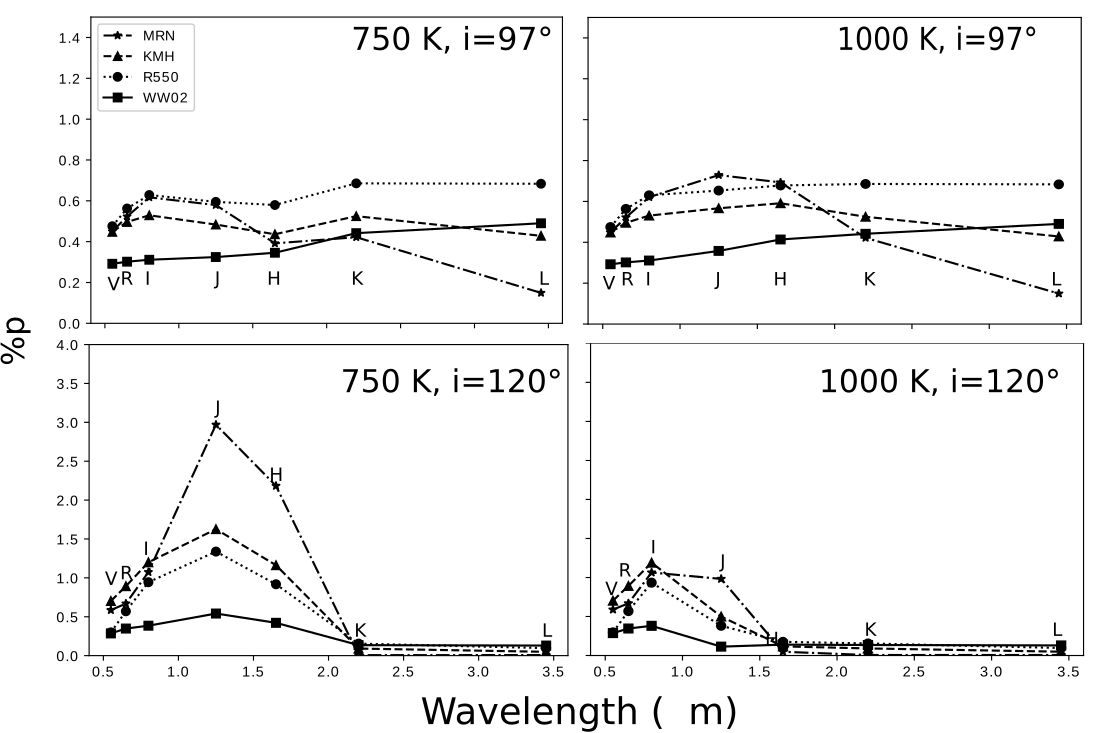}
\caption{Simulated polarization as a function of wavelength for emitting dust at two viewing angles, two temperatures, and a reference optical depth of $\tau_0 = 2.0$. Different lines and symbols denote different dust types. Error bars are smaller than the plotted points.}
\label{sed_diffdust_emission}
\end{figure*}

In Fig~\ref{sed_diffdust_emission}, we display the pSEDs for the higher optical depth case. At $i=97\degr$, we see very little difference in the behaviour of the pSEDs between temperatures for same dust type. The polarization behaviour in this case is very similar to the lower optical depth, no-emission scenario (Fig.~\ref{sed_diffdust}, top left panel). This is because at $i \sim90\degr$, single scattering is the dominant source of polarization. Thus, even when the dust emits a larger fraction of the total photons, the polarization at this angle does not change.

At $i=120\degr$, the strong temperature dependence discussed in \S~\ref{result_emission_ur_temp} manifests in the sharp decrease in polarization from 750 K to 1000 K. In addition, all dust models show a significant peak in polarization at near-IR wavelengths ($IJ$ bands). This peak is present for both temperatures we tested, but is stronger at the lower temperature of 750 K and for the ISM dust types (MRN and KMH). Thus, the $IJ$ polarization slope is likely to be a good observational dust type diagnostic at this viewing angle (subject to the ISP considerations discussed above). 

\subsection{Dust between the star and bow shock}\label{dust_in}

In a more physical scenario, there will be dust and gas between the star and the bow shock. In the case of a cool central star, dust may be present in most of the region between the star and bow shock, whereas for a hotter star there will be an empty volume around the star up to the dust condensation radius. We modeled the cool-star scenario, which serves as a limiting case compared to our previous models with no dust between the bow shock and the star. As the stellar temperature increases, we expect the dust-free interior region to increase in size and the polarization behaviour to approach that of our previously discussed results.

\begin{figure}
0\includegraphics[width=\linewidth]{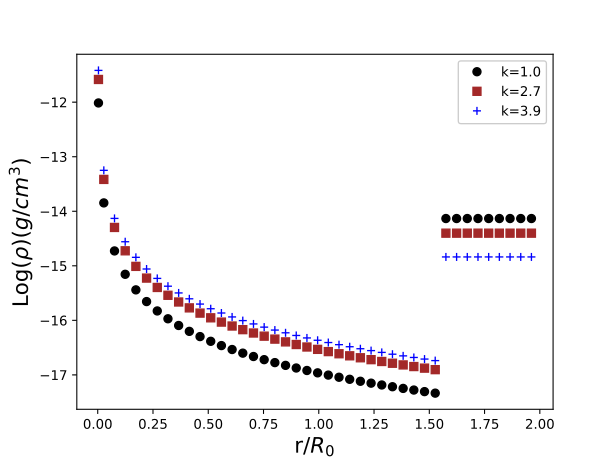}
\caption{Density variation along the line of sight at a representative viewing angle of $90\degr$, for the case of dust within the bow shock with a reference optical depth $\tau_0=2$. 
Different symbols listed in the legend represent different values of the code's density scale factor $k$ (\S~\ref{dust_in}).}
\label{densfunc}
\end{figure}

To simulate this case, as a first approximation, we added interior dust with a radially decaying density. We used the functional form $\rho = k \rho_0 (r/R_0)^{-2}$, where $r$ is the distance from the star, $\rho_0$ is the reference density at the standoff radius $R_0$, and $k$ is a scale factor of order unity that facilitates changing the density within the code. We chose this density function to be similar to recent smooth particle hydrodynamic (SPH) models of Betelgeuse's bow shock \citep{Mohamed_2012}. We set the reference density at $\rho_0=1.09\times10^{-17}$ g cm$^{-3}$, such that the interior dust density was 1--3 orders of magnitude lower than the densities within the bow shock we used in previous models. Fig.~\ref{densfunc} depicts the line-of-sight density at a representative viewing angle of $90\degr$ for several values of $k$. The optical depth of the system at a given viewing angle now takes into account the interior dust density, so for a given reference optical depth $\tau_0$, the density in the bow shock changes slightly as a function of $k$. When $k=0$, we recover the case of no material within the shock.

\begin{figure*}
\includegraphics[width=\textwidth]{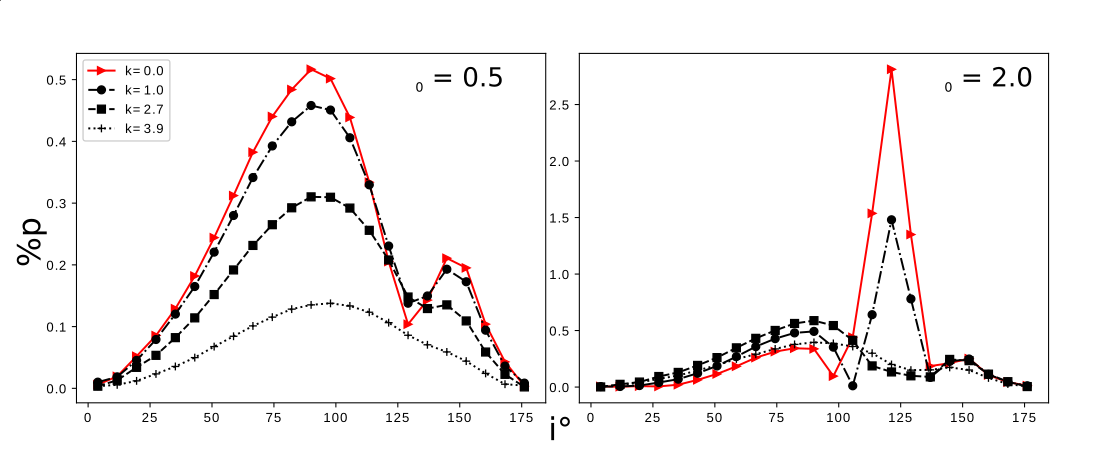}
\caption{Simulated polarization as a function of viewing angle for the case of dust between the star and the bow shock. The curves represent different values of the code's scale factor $k$, with $k=0$ corresponding to no interior dust (\S~\ref{dust_in}).  All of the models shown here are for KMH dust at 2.2 $\mu$m ($K$ band).}
\label{pva_dusntin}
\end{figure*}

Although \textit{SLIP} allows us to choose different scattering laws for the bow shock and the interior, for simplicity we chose KMH dust for both in these tests. We also considered only the case in which the dust itself does not emit. The results for the $K$ band are shown in Fig.~\ref{pva_dusntin}. We found that including dust between the star and the bow shock suppresses the overall polarization level for all optical depths (Fig.~\ref{pva_dusntin} shows only two representative $\tau_0$ values), consistent with the model results of \citet{Shahzamanian_2016} and \citet{Zajacek_2017}. This is because when dust fills the cavity, photons illuminate the shock from a wide range of directions rather than radially from the star. This removes any preferential scattering planes and results in lower polarization. 
The secondary polarization peak near $120\degr$ is suppressed even more than near $90\degr$ peak because photons must travel farther within the interior dust to escape at the higher angle; multiple scattering along a longer path more effectively removes the ``memory" of their origin at the star. Cases with hot central stars are not shown here, but because they have less interior dust, they should produce polarization values intermediate between the cool-star case and the case with no dust.

The presence of dust inside the shock introduces degeneracy into the model results we have presented so far by decreasing the predicted polarization for a given set of bow shock properties. Our initial tests show that even the ratio between polarization values at different wavelengths does not always remain constant when interior dust of the same type is added. This degeneracy makes it difficult to draw general conclusions regarding observational strategy; the depolarizing effect of interior dust must be recognized as a caveat to interpreting observed polarimetric results. In addition, this case is unlike our previously presented simulations in that the exact value of the standoff radius $R_0$ affects the observed polarization by controlling the optical depth of the interior dust. Nevertheless, individual objects can still be interpreted reliably with the methods we present here. Constraints from supporting observations will allow us to construct models focused more closely on particular objects and thereby draw meaningful conclusions on a case-by-case basis.

\section{Observational implications}
\label{comparison_observation}
\subsection{Resolved bow shock}
\label{sec:obs_r}
\begin{figure}
    \centering
    \includegraphics[width=\columnwidth]{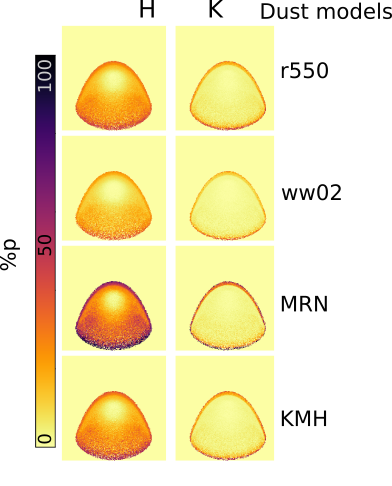}
    \caption{Simulated polarization maps for a resolved model bow shock with emitting dust at 750 K and a reference optical depth of $\tau_0 = 0.5$, viewed at an inclination angle of $125\degr$. These are good comparisons to the resolved bow shock source IRS~1W (\citealt{Buchholz_2011}; \S~\ref{sec:obs_r}).}
    \label{fig:irs1w}
\end{figure}

Although polarization observations of resolved bow shocks are rare in the literature, our results show that they are potentially very useful diagnostic tools. Comparing observed maps in polarization and polarized intensity with our simulated images will allow an observer to infer the properties of the dust in the bow shock. Polarized intensity maps can constrain the inclination angle for cases of higher optical depth (\S~\ref{result_noemission_r_opt}; Fig.~\ref{map_othick}). Observing the spatial variation of the polarization of a resolved bow shock at $J$ or $H$ band can diagnose the temperature in the shock region (\S~\ref{result_emission_r_temp}; Fig~\ref{pmap_emission}), while maps in the $H$ or $K$ band also provide information about the size of the dust particles creating the polarization (\S~\ref{dusttype_r_emission}; Fig.~\ref{pmap_emission}).

\citet{Rauch_2013} presented intrinsic (ISP-corrected) $K_\textrm{S}$-band polarimetric observations of a resolved stellar wind bow shock around IRS 8. Although they obtained only 9 polarization measurements across the bow shock, we can make a few comparisons with our resolved simulations. These authors found a decrease in polarization at the apex of the IRS 8 bow shock compared with the outer edges and very low polarization in the center of the region. In our $K$-band models, this behaviour would suggest the presence of thermally emitting dust at a relatively high optical depth (Fig.~\ref{pmap_emission}, lower panels; we note that \citeauthor{Rauch_2013} constrained the inclination angle of the IRS 8 bow shock to $i\sim50\degr$, which agrees well with the representative results in this figure). This is consistent with the authors' inference of high particle density in the apex, causing a decrease in polarization due to multiple scattering effects. 

Since polarization behaviour is independent of standoff radius in our simulations (assuming no dust inside the bow shock), we can utilize the standoff radius of IRS 8 from \citet[][$R_0 = 3.5 \times 10^{16}$ cm]{Rauch_2013} and the optical depth of $\tau_0 = 2.0$ from our simulations to calculate the local ISM density. We begin with the definition of optical depth: $\tau = \rho \kappa \Delta R$. From Eq.~\ref{rho}, we can write $\rho\Delta R = CR_0\rho_I$ and thus $\tau=C\kappa R_0\rho_I$, where $\rho_I$ is the ISM density we wish to find and $C$ is constant for a given viewing angle and $\alpha$ value. In this case, $C\approx1.7$. For KMH dust in the $K$ band, $\kappa = 22.52$ cm$^2$/g, and in our models, when $\tau_0 = 2.0$, the optical depth at $50\degr$ is 1.1 (Fig.~\ref{emission}). With these values, we find $\rho_I=8.2\times 10^{-19}$ g/cm$^3$ for IRS 8. The ISM number density is given by $n = \frac{\rho_I}{\tilde{m}}$, where $\tilde{m} = 2.17 \times 10^{-24}$ g for solar abundances. With this expression, we find the local ISM number density for IRS 8 to be $\sim10^5$ cm$^{-3}$. This value is higher than the  value of $10^4$ cm$^{-3}$ assumed for the ISM near IRS 8 by \citep{Rauch_2013}, but similar to the estimate of $10^5$ cm$^{-3}$ found by \citet{Tanner_2002}. Discrepancies between our models and other estimates could be due to the presence of dust between the source and bow shock, or to the dust grains being nonspherical. 

In fact, \citeauthor{Rauch_2013} found on-sky position angles of $\sim0\degr$ at all their measured locations, corresponding to orientations roughly perpendicular to the shock at the bow head. This behavior of the position angle is significantly different than that produced by our models (\S~\ref{results}),
which suggests that the dust in IRS 8 is not similar to any of the spherical-grain types we consider. The authors note that their measured angle agrees well with the direction of the magnetic field in the Galactic northern arm, and conclude that the dust grains in IRS 8 are aspherical and aligned along this field. Additional modeling taking into account interior dust and grain alignment \citep[e.g.,][]{Whitney_2002} would be required to reproduce these polarimetric observations.

\citet{Buchholz_2011,buchholz2013} obtained intrinsic polarization maps of bow shocks around the sources IRS~21  and IRS~1W in the Galactic Center in $H$, $K_\textrm{S}$, and $L$p bands. IRS~21 displays a predominantly round shape -- \citet{Buchholz_2011} deem it a ``marginally resolved bow shock" -- so this source is not a good comparison with our models. However, IRS~1W has a well defined bow shock shape and high-quality polarization maps in 3 wavebands. We compared the maps of IRS~1W (Figs. 18 and 19 in \citealt{Buchholz_2011}; Fig. 13 in \citealt{buchholz2013}) with our models, assuming a 125\degr~viewing angle is a good approximation for this object ($i=137\degr$; \citealt{Sanchez-Bermudez_2014}). \citet{Buchholz_2011,buchholz2013} found intrinsic polarization magnitudes ranging between $10-20\%$ for IRS~1W, with generally higher values at $H$ than at $K_\textrm{S}$ and $L$p. Their maps also showed that the polarization at the bow head is lower than in the wings at $K_\textrm{S}$ and $L$p, but not at $H$.
Our models without dust emission (Figs.~\ref{map_othin} and~\ref{map_othick}) show little difference between $H$ and $K$ bands for any dust type, so we conclude the dust in this system is warm enough to contribute significant emission. Although our results in Fig.~\ref{pmap_emission} only show $i=55\degr$, we can infer from them that the dust temperature is likely closer to 750 K than 1000 K, because only at 750 K do the $H$ band maps show significantly higher polarization than the $K$ band maps. (\citealt{moultaka2004} estimated 900 K for this source via blackbody fitting.) We can also constrain the optical depth to be small, because at $\tau_0=2$, our $K$-band maps show nearly zero polarization in the center of the bow shock, which is inconsistent with \citeauthor{Buchholz_2011}'s observations. 

Fig.~\ref{fig:irs1w} shows our model results for $i=125\degr$, $T=750$ K, and $\tau_0=0.5$; these are our best comparisons to IRS~1W. All the dust types we consider show larger polarization at $H$ than at $K$, but the R550 dust type produces the clearest $K$-band decrease from the wings to the bow head. It also matches the results of  \citet{Buchholz_2011} well in terms of polarization magnitude, but this is not a strong constraint, as dust inside the shock decreases the outgoing polarization (\S~\ref{dust_in}). However, our simulated position angles produce a centrosymmetric pattern even in cases of dust emission, whereas \citet{Buchholz_2011,buchholz2013} found a consistent polarization orientation of $\Phi\sim-75\degr$ across the bow shock (parallel to the bow head, in contrast with the results of \citealt{Rauch_2013}). Thus, while the spherical dust grains we consider could produce the behaviour of the polarization magnitudes observed in IRS~1W, the position angle behaviour suggests elongated grains are more likely, as discussed in \citet{Buchholz_2011}. With this caveat in mind, we can calculate the local ISM number density around IRS~1W using the same technique as above for IRS 8, with $R_0=689$ AU, $i=125\degr$, $\tau_0=0.5$, and $C\approx13$. We find $n\sim7\times 10^4$ cm$^{-3}$ for this case, which is similar to the predicted values of $10^4-10^5$ cm$^{-3}$ found by \citet{Rauch_2013} and \citet{Tanner_2002}, respectively. Constraining the dust type and other quantities more tightly would require a more detailed comparison of the observational images with model results, taking elongated grains into account and including various amounts of dust inside the shock.

\subsection{Unresolved bow shock}
\label{sec:obs_ur}
In cases in which the system is unresolved but a bow shock is suspected due to a star's high velocity, a typical observation will yield a single polarization value at a given wavelength. Such a measurement can be compared with our unresolved simulation results, such as those in Fig.~\ref{electroncomp} or Fig.~\ref{pva_dusntin}, to place limits on the optical depth and inclination angle of the object. Doing this reliably requires estimating the ISP contribution to the measurement. Additional modeling may be necessary for individual objects to constrain the contribution of dust inside the bow shock.

However, our results demonstrate that we can significantly improve the diagnostic power of polarimetry for an unresolved source by obtaining multiple polarization measurements in different wavebands to create a pSED. 
While the morphology of these pSEDs is affected by the ISP, this contribution can be estimated and removed in the usual ways or added to simulation results to expand their predictive ability. The shapes of the pSEDs in our models are insensitive to the presence of interior dust at least at some viewing angles (\S~\ref{dust_in}). Comparing an observed pSED with  Figs.~\ref{sed_diffdust}, \ref{sed_difftemp}, and \ref{sed_diffdust_emission} can provide important information about the dust making up the bow shock, including its grain size, composition, and temperature (if emitting). In many cases, especially when constraints on viewing angle and ISP can be inferred from other observations, it is not necessary to construct a full 7-band pSED such as we display here; strategic observations at 2--3 wavebands  provide pSED slopes with the power to distinguish among dust types and constrain the dust temperature.

Several authors have obtained unresolved polarization  observations of bow shocks that we can compare with our simulations. \citet{Lin_2019} found the intrinsic (ISP-corrected) broadband optical polarization of the candidate bow shock system HD 230561 (spectral type A5) to be ~1.05\%. Comparing this observation with our model for KMH dust in the $V$-band (assuming no dust inside the shock) allowed these authors to constrain the viewing angle to $i\approx60{\degr}$ or $80{\degr}\lesssim i \lesssim130{\degr}$ and the optical depth of the shock region to $\tau_0\gtrsim2$. An additional $K$-band observation of this object would allow more robust conclusions.
 
\cite{Neilson_2014} proposed that an unresolved stellar wind bow shock could explain the large $V$-band intrinsic polarization ($p_V\approx 1-6\%$) in the semi-regular variable V CVn. These authors attributed the time variability of their polarization measurements to pulsation-related variability in the density of the stellar wind. This explanation is consistent with our models with dust inside the bow shock, which show that the degree of polarization can vary significantly depending on the amount of interior dust (Fig.~\ref{pva_dusntin}). Even without interior dust, the maximum unresolved $V$ polarization produced by the models we have presented is only $\sim1\%$ (for $\tau_0=2.0$, high viewing angle, and ISM-type dust; Figs.~\ref{electroncomp},~\ref{sed_diffdust},~\ref{sed_difftemp},~\ref{sed_diffdust_emission}). Given our general finding that higher optical depths in the bow shock result in higher polarization (\S~\ref{results}), this suggests that the shock region in V CVn may itself have a high density. The large polarization produced by this dense shock could then be diluted to varying extents depending on the amount of interior dust. More detailed modeling of this object is clearly warranted.

In addition to the resolved observations discussed in \S~\ref{sec:obs_r}, \cite{Buchholz_2011,buchholz2013} obtained integrated $HK_\textrm{s}L$p polarization measurements of many Galactic Center sources, including the  known bow shock systems IRS~1W, IRS 5, IRS 10W, and IRS~21. For IRS~1W, they measured intrinsic (ISP-corrected) total polarization values of $P_H=6.9\%$, $P_{K_\textrm{S}}=7.8\%$, and $P_{L\textrm{p}}=8.9\%$ with uncertainties of order $0.5\%$. (While the polarization magnitude is higher overall in the resolved $H$-band images than at the other bands, position angle effects cause the integrated value to decrease below the $K_\textrm{S}$- and $L$p-band polarization in the unresolved case.) In our simulated pSEDs at viewing angles near $120\degr$, this increase in unresolved polarization from $H$ to $K$ occurs only for the case of no dust emission or dust temperatures of 500 K, and never occurs for MRN dust (Fig.~\ref{sed_diffdust}; Fig.~\ref{sed_difftemp}). 

All our simulations also show an increase in polarization from $K$ to $L$. However, none of the models we present here produces the high polarization magnitudes observed by \citet{Buchholz_2011,buchholz2013}. This may indicate that the optical depths are high in this bow shock (\S~\ref{results}) or that the dust grains in the system are elongated, as we discuss in \S~\ref{sec:obs_r}. However, we can test whether our simulated pSEDs show a similar morphology to the observed ones. \citet{buchholz2013} report an $HK_\textrm{s}L$p ratio of 0.9:1:1.1 for IRS~1W. Assuming $\tau_0=2$ and $i=120\degr$, we find the closest match for KMH dust with no emission (0.96:1:1.1). KMH dust with emission at $750$~K produces a much less satisfactory match (3.9:1:1.05). These ratios may, however, be quite different for the case of elongated aligned grains.

IRS 5 shows lower near-IR polarization values more comparable with our model results  ($P_{K_\textrm{S}}=2.1\%$ and $P_{L\textrm{p}}=2.5\%$; \citealt{buchholz2013}). For an approximate inclination angle of $i=120\degr$ (\citealt{Sanchez-Bermudez_2014} estimated $i=130\degr$), we find reasonable matches with our simulations for non-emitting KMH dust with $\tau_0=2$ (Fig.~\ref{sed_diffdust}). WW02 and R550 dust also produce increased polarization from $K$ to $L$, but the observed $p_L/p_K$ ratio of 1.2 matches best with our simulated KMH ratio ($p_L/p_K=1.1$). Although our simulated polarization magnitudes are slightly larger than the observed ones ($p_K$=2.8\%, $p_L=3.1\%$), the presence of dust inside the bow shock could easily reduce these values to the observed ones. We conclude that spherical grains of KMH dust could produce the observed magnitude of the near-IR polarization in IRS~5. However, our unresolved simulations all produce net polarization close to parallel to the axis of the bow shock \citepalias[see also][]{Shrestha_2018}, whereas the observed position angle of IRS~5 is perpendicular to the axis \citep{buchholz2013}. Thus, it is still most likely that aligned grains produce the polarization in this case.

The nearby sources IRS~10W and IRS~21 show intrinsic polarization values significantly higher than any of our unresolved polarization models ($P_{K_\textrm{S}}=4.2\%$ and $P_{L\textrm{p}}=5.6\%$ for IRS~10W, $P_{K_\textrm{S}}=6.1\%$ and $P_{L\textrm{p}}=15.0\%$ for IRS~21; \citealt{Buchholz_2011,buchholz2013}). As in the case of IRS~1W, these large values likely point to very dense dust or aligned dust grains.

\citet{Shahzamanian_2016} and \citet{Zajacek_2017} used a MCRT code to model the polarization behaviour of the Dusty S-cluster object (DSO/G2), which has a very large polarization of $P_{K_\textrm{S}}>20\%$. Using KMH-like dust and a bow shock geometry with constant density in the shock layer, they found a maximum $\%p_{K_\textrm{S}}=4.1\%$, with resolved polarization and position angle behavior similar to those we present here (their Fig.~13, top row; we note especially the roughly centrosymmetric position angle behavior of these position angle maps). These authors concluded that additional scattering structures must create the additional polarization in the DSO. The models we present here achieve the highest polarization signal ($\sim3\%$) at $K$-band for KMH dust at high optical depth of $\tau_0=2.0$ and inclination angle of $120\degr$. However, at higher reference optical depths ($\tau_0 > 4$), our simulations can produce polarization of over $10\%$ at inclination angles $90\degr$ and higher. \citet{Zajacek_2017} found similar large polarization values at high optical depth (M. Zaja\v cek, priv. comm.), but were motivated to add other dust structures because the bow shock layer is not expected to be so dense in the  Galactic center environment \citep{Scoville_2013}.

In summary, the multiwavelength polarization simulation results we present here can reasonably reproduce the polarization of some bow shock systems like HD 230561 \citep{Lin_2019}. Because we have considered spherical grains only, our simulations are less applicable to cases where magnetic fields likely play a role in aligning dust grains, such as the Galactic Center sources studied by \citet{Buchholz_2011,buchholz2013}, but they may still provide useful insights. For systems with large polarization magnitudes such as V CVn \citep{Neilson_2014} or the DSO \citep{Shahzamanian_2016,Zajacek_2017}, our simulations provide lower limits to the observed polarization that can guide future modeling by diagnosing the presence of more complex circumstellar structures. In all cases, we find the most useful constraints are possible with polarization observations at multiple wavelengths, and encourage observers to make this a standard practice.

\section{Conclusions and future work}\label{conclusions}
 We used a Monte Carlo radiative transfer code, \textit{SLIP} (\citealt{Hoffman_2007}; \citetalias{Shrestha_2018}), to investigate the polarization arising from dust scattering within a stellar wind bow shock defined by the analytical model of \citet{Wilkin_1996}. We studied how various parameters affect the polarization behaviour for both resolved and unresolved bow shocks. The major conclusions are presented here:

\begin{enumerate}

    \item We find that the resulting polarization is highly dependent on the inclination angle, wavelength, and dust grain properties. 
   
 In general, the polarization produced by dust scattering behaves qualitatively similarly to that produced by electron scattering \citetalias{Shrestha_2018}, but with lower polarization magnitudes for similar simulation parameters. As a function of inclination angle, the dust scattering models at low optical depths  produce polarization resembling the analytical results derived by \citet[][modified by absorption effects]{brown_1977}, but the peak of the primary polarization curve is shifted to larger angles than $i=90\degr$. As in the electron-scattering case, higher optical depths produce a second polarization peak near $i=130\degr$, corresponding to a sign flip of the Stokes $q$ parameter. Overall, of the dust types we tested, MRN dust at 2.2 $\micron$~ produces the most similar results to electron scattering.
 
 \item We studied how the dust parameters in the analytical Henyey-Greenstein function affect the polarization behavior of simulations using this function to describe dust scattering. We find that the simulated polarization increases with increasing albedo $a$  and decreases with increasing scattering asymmetry $g$.
 
 \item We simulated the polarization of dusty bow shocks at different wavelengths (\textit{VRIJHKL} bands). When the dust in the bow shock does not emit, the polarization behavior is dominated by the optical depth and the dust type, and is relatively independent of dust temperature. For resolved cases at low optical depth, the resulting polarization maps depend strongly on inclination angle, while the polarized intensity maps show little variation. By contrast, at higher optical depth, both polarized intensity and polarization degree show spatial distributions that vary with inclination angle. The maps we created show minor variation with wavelength, but primarily in the amount of polarization/polarized intensity rather than the morphology. 
 
 For unresolved cases, the variation of polarization with wavelength depends strongly on the dust type. Dust types with larger grains (WW02 and R550) produce polarization that increases with wavelength, whereas ISM-like dust with smaller grains (KMH and MRN) shows Serkowski-like behaviour for lower optical depths \citep{Serkowski_1975}. KMH dust diverges from this behaviour for larger optical depths and higher inclination angles. The model pSEDs we created can serve as useful comparisons with multiwavelength polarization observations, allowing some system properties to be constrained.
 
 \item  When the dust in the bow shock emits its own photons, the resulting polarization depends on the dust temperature in addition to optical depth and dust type. Our resolved maps show that this polarization varies with wavelength for a given  temperature. 
 At longer wavelengths, the contribution of the dust emission becomes particularly important due to the shifting of the Planck peak; the polarization maps at these wavelengths thus resemble those of the ``distributed emission" case for electron scattering \citetalias{Shrestha_2018}.
 
 In the lower optical depth regime for unresolved bow shocks, dust emission does not affect the behaviour of polarization with wavelength. For higher optical depths, dust emission from the bow shock combined with multiple scattering effects creates pSEDs with distinct temperature- and inclination-dependent morphologies, particularly at wavelengths longer than $I$ band. Thus, multiband polarization observations at these longer wavelengths provide information about the dust temperature if the inclination angle can be estimated.
 
Similarly, at low optical depths the dust type does not meaningfully change the observed pSEDs for unresolved bow shock sources. However, at higher optical depths and longer wavelengths, different dust types produce distinct pSED morphologies that can be sampled with polarization observations.
 
 \item We also investigated the polarization behaviour of our simulations when dust fills the interior of bow shock, using the KMH dust model as a test case. The existence of interior dust suppresses the resulting polarization near both $90\degr$ and $120\degr$ regardless of the optical depth of the bow shock. 
\end{enumerate}

When interpreting these results, we need to keep in mind the several simplifying assumptions we made in these models. We chose a specific standoff radius $R_0$ (\S~\ref{method}) for all the models presented here. We tested the effect of changing $R_0$ on the polarization behaviour and found it has no significant impact except in the case of dust interior to the bow shock (\S~\ref{dust_in}). More detailed modeling would be required to interpret such cases. We also fixed the parameter $\alpha={V_*}/{V_w}=0.1$, which is used to calculate the density of the bow shock (Section~\ref{method}). This parameter would need to be varied or assumed for the specific modeled object in order to extract information about the speed of the stellar wind. 

In creating our simulation results, we have made no attempt to account for additional polarization signal arising from interstellar dust. ISP contributions add vectorially with the intrinsic polarization signal, so it is difficult to predict their effect \textit{a priori}. However, several well known methods exist to estimate or constrain the ISP for a given source. Our results suggest that obtaining polarization observations of bow shock sources in multiple wavebands is the best way to constrain system parameters, as this allows comparison of the pSED slope with both intrinsic predictions and ISP models. We are obtaining multiwavelength observations of a new sample of stellar wind bow shocks and will investigate their properties via comparisons with our model grid, taking ISP contributions into account.

We have also not considered the contribution of localized dust external to the bow shock structure \citep{Meyer_2014,Henney_2019}; such a scenario may arise if stellar wind dust penetrates outward into the local ISM, as shown by \citet{Marle_2011}. We ran  preliminary \textit{SLIP} simulations adding an outer dust layer with the same shape and dust composition as the bow shock and a constant, lower density. This produced a slight ($< 0.1\%$) increase in polarization, most likely due to the additional optical depth contributed by the exterior dust. In future work, we plan to model this region in more detail, investigating different dust types from the bow shock and more complex density functions.

It is important to note that the analytic bow shock shape and density we have adapted from \citet{Wilkin_1996} assumes a stable and highly evolved bow shock, as shown by \citet{Mohamed_2012} using their smooth particle hydrodynamic models. Thus, resolved images from our simulations represent that evolved state, and may not provide good comparisons with younger bow shocks. However, the simulated images could  be compared with observations to assess the evolutionary phase of the target bow shock. In other cases with prominent instabilities \citep{Meyer_2014,Meyer_2015}, the bow shock will exhibit a different overall structure. We plan to investigate these more complex bow shock morphologies in a future contribution. 

In a realistic bow shock scenario, both electrons and dust may scatter light and produce polarization signatures. In principle, \textit{SLIP} can treat a case where both types of scattering exist, even if they occur in the same spatial region. This would require a more detailed definition of the circumstellar environment than we have considered in these initial models, but we plan to investigate the combined effects of the two types of scattering in future simulations of specific bow shock systems. 

\section*{Acknowledgments}
This work has been supported by the National Science Foundation (AST-1210372 to JLH) and by a Sigma Xi Grant-in-Aid of Research to MS. We thank B. Whitney for productive discussions and for sharing her dust scattering data files, as well as M. Zaja{\v c}ek for sharing his code and results and the anonymous referee for many helpful comments and suggestions. This work used the Extreme Science and Engineering Discovery Environment (XSEDE), which is supported by the National Science Foundation under award ACI-1548562. We used their Stampede2 cluster to run some of our simulations, under allocation ID TG-AST120067. The plots were made using the Matplotlib Python package \citep{Hunter:2007}. We acknowledge that most of the work for this paper was done on Cheyenne and Arapaho lands.

\section*{Data availability}
The data used in this article can be shared on reasonable request to the corresponding author.






\newpage
 
\bibliographystyle{mnras}
\bibliography{dust}

\begin{thebibliography}{}
\makeatletter
\relax
\def\mn@urlcharsother{\let\do\@makeother \do\$\do\&\do\#\do\^\do\_\do\%\do\~}
\def\mn@doi{\begingroup\mn@urlcharsother \@ifnextchar [ {\mn@doi@}
  {\mn@doi@[]}}
\def\mn@doi@[#1]#2{\def\@tempa{#1}\ifx\@tempa\@empty \href
  {http://dx.doi.org/#2} {doi:#2}\else \href {http://dx.doi.org/#2} {#1}\fi
  \endgroup}
\def\mn@eprint#1#2{\mn@eprint@#1:#2::\@nil}
\def\mn@eprint@arXiv#1{\href {http://arxiv.org/abs/#1} {{\tt arXiv:#1}}}
\def\mn@eprint@dblp#1{\href {http://dblp.uni-trier.de/rec/bibtex/#1.xml}
  {dblp:#1}}
\def\mn@eprint@#1:#2:#3:#4\@nil{\def\@tempa {#1}\def\@tempb {#2}\def\@tempc
  {#3}\ifx \@tempc \@empty \let \@tempc \@tempb \let \@tempb \@tempa \fi \ifx
  \@tempb \@empty \def\@tempb {arXiv}\fi \@ifundefined
  {mn@eprint@\@tempb}{\@tempb:\@tempc}{\expandafter \expandafter \csname
  mn@eprint@\@tempb\endcsname \expandafter{\@tempc}}}

\bibitem[\protect\citeauthoryear{{Brown} \& {McLean}}{{Brown} \&
  {McLean}}{1977}]{brown_1977}
{Brown} J.~C.,  {McLean} I.~S.,  1977, \aap, \href
  {http://cdsads.u-strasbg.fr/abs/1977A26A....57..141B} {57, 141}

\bibitem[\protect\citeauthoryear{{Buchholz}, {Witzel}, {Sch{\"o}del}, {Eckart},
  {Bremer}  \& {Mu{\v z}i{\'c}}}{{Buchholz} et~al.}{2011}]{Buchholz_2011}
{Buchholz} R.~M.,  {Witzel} G.,  {Sch{\"o}del} R.,  {Eckart} A.,  {Bremer} M.,
   {Mu{\v z}i{\'c}} K.,  2011, \mn@doi [\aap] {10.1051/0004-6361/201117300},
  \href {http://adsabs.harvard.edu/abs/2011A26A...534A.117B} {534, A117}

\bibitem[\protect\citeauthoryear{{Buchholz}, {Witzel}, {Sch{\"o}del}, {Eckart},
  {Bremer}  \& {Muzic}}{{Buchholz} et~al.}{2012}]{Buchholz_2012}
{Buchholz} R.~M.,  {Witzel} G.,  {Sch{\"o}del} R.,  {Eckart} A.,  {Bremer} M.,
   {Muzic} K.,  2012, in Journal of Physics Conference Series. p. 012021,
  \mn@doi{10.1088/1742-6596/372/1/012021}

\bibitem[\protect\citeauthoryear{{Buchholz}, {Witzel}, {Sch{\"o}del}  \&
  {Eckart}}{{Buchholz} et~al.}{2013}]{buchholz2013}
{Buchholz} R.~M.,  {Witzel} G.,  {Sch{\"o}del} R.,   {Eckart} A.,  2013, \aap,
  \href {https://ui.adsabs.harvard.edu/abs/2013A&A...557A..82B} {557, A82}

\bibitem[\protect\citeauthoryear{{Chandrasekhar}}{{Chandrasekhar}}{1960}]{Chandrasekhar_1960}
{Chandrasekhar} S.,  1960, {Radiative transfer}

\bibitem[\protect\citeauthoryear{{Clayton}, {Wolff}, {Sofia}, {Gordon}  \&
  {Misselt}}{{Clayton} et~al.}{2003}]{Clayton_2003}
{Clayton} G.~C.,  {Wolff} M.~J.,  {Sofia} U.~J.,  {Gordon} K.~D.,   {Misselt}
  K.~A.,  2003, \mn@doi [\apj] {10.1086/374316}, \href
  {http://adsabs.harvard.edu/abs/2003ApJ...588..871C} {588, 871}

\bibitem[\protect\citeauthoryear{{Code} \& {Whitney}}{{Code} \&
  {Whitney}}{1995}]{Code_1995}
{Code} A.~D.,  {Whitney} B.~A.,  1995, \mn@doi [\apj] {10.1086/175363}, \href
  {http://adsabs.harvard.edu/abs/1995ApJ...441..400C} {441, 400}

\bibitem[\protect\citeauthoryear{{Cotera} et~al.,}{{Cotera}
  et~al.}{2001}]{Cotera_2001}
{Cotera} A.~S.,  et~al., 2001, \mn@doi [\apj] {10.1086/321627}, \href
  {http://adsabs.harvard.edu/abs/2001ApJ...556..958C} {556, 958}

\bibitem[\protect\citeauthoryear{{Henney} \& {Arthur}}{{Henney} \&
  {Arthur}}{2019a}]{Henney_2019}
{Henney} W.~J.,  {Arthur} S.~J.,  2019a, \mn@doi [\mnras]
  {10.1093/mnras/stz1043}, \href
  {https://ui.adsabs.harvard.edu/abs/2019MNRAS.486.3423H} {486, 3423}

\bibitem[\protect\citeauthoryear{{Henney} \& {Arthur}}{{Henney} \&
  {Arthur}}{2019b}]{Henney_2019b}
{Henney} W.~J.,  {Arthur} S.~J.,  2019b, \mn@doi [\mnras]
  {10.1093/mnras/stz1130}, \href
  {https://ui.adsabs.harvard.edu/abs/2019MNRAS.486.4423H} {486, 4423}

\bibitem[\protect\citeauthoryear{{Henyey} \& {Greenstein}}{{Henyey} \&
  {Greenstein}}{1941}]{Henyey_1941}
{Henyey} L.~G.,  {Greenstein} J.~L.,  1941, \mn@doi [\apj] {10.1086/144246},
  \href {https://ui.adsabs.harvard.edu/abs/1941ApJ....93...70H} {93, 70}

\bibitem[\protect\citeauthoryear{{Hoffman}}{{Hoffman}}{2007}]{Hoffman_2007}
{Hoffman} J.~L.,  2007, in Revista Mexicana de Astronomia y Astrofisica
  Conference Series. pp 57--63 (\mn@eprint {} {astro-ph/0612244})

\bibitem[\protect\citeauthoryear{Hunter}{Hunter}{2007}]{Hunter:2007}
Hunter J.~D.,  2007, \mn@doi [Computing in Science \& Engineering]
  {10.1109/MCSE.2007.55}, 9, 90

\bibitem[\protect\citeauthoryear{{Jayasinghe} et~al.,}{{Jayasinghe}
  et~al.}{2019}]{Jayasinghe_2019}
{Jayasinghe} T.,  et~al., 2019, \mn@doi [\mnras] {10.1093/mnras/stz1738}, \href
  {https://ui.adsabs.harvard.edu/abs/2019MNRAS.tmp.1691J} {p.~1691}

\bibitem[\protect\citeauthoryear{{Johnson} \& {Morgan}}{{Johnson} \&
  {Morgan}}{1953}]{Johnson_1953}
{Johnson} H.~L.,  {Morgan} W.~W.,  1953, \mn@doi [\apj] {10.1086/145697}, \href
  {https://ui.adsabs.harvard.edu/abs/1953ApJ...117..313J} {117, 313}

\bibitem[\protect\citeauthoryear{{Kim}, {Martin}  \& {Hendry}}{{Kim}
  et~al.}{1994}]{Kim_1994}
{Kim} S.-H.,  {Martin} P.~G.,   {Hendry} P.~D.,  1994, \mn@doi [\apj]
  {10.1086/173714}, \href {http://adsabs.harvard.edu/abs/1994ApJ...422..164K}
  {422, 164}

\bibitem[\protect\citeauthoryear{{Kobulnicky} et~al.,}{{Kobulnicky}
  et~al.}{2016}]{Kobulnicky_2016}
{Kobulnicky} H.~A.,  et~al., 2016, \mn@doi [\apjs]
  {10.3847/0067-0049/227/2/18}, \href
  {http://adsabs.harvard.edu/abs/2016ApJS..227...18K} {227, 18}

\bibitem[\protect\citeauthoryear{{Kobulnicky}, {Chick}  \&
  {Povich}}{{Kobulnicky} et~al.}{2018}]{Kobulnicky_2018}
{Kobulnicky} H.~A.,  {Chick} W.~T.,   {Povich} M.~S.,  2018, \mn@doi [\apj]
  {10.3847/1538-4357/aab3e0}, \href
  {https://ui.adsabs.harvard.edu/abs/2018ApJ...856...74K} {856, 74}

\bibitem[\protect\citeauthoryear{{Kobulnicky}, {Chick}  \&
  {Povich}}{{Kobulnicky} et~al.}{2019}]{Kobulnicky_2019}
{Kobulnicky} H.~A.,  {Chick} W.~T.,   {Povich} M.~S.,  2019, \mn@doi [\aj]
  {10.3847/1538-3881/ab2716}, \href
  {https://ui.adsabs.harvard.edu/abs/2019AJ....158...73K} {158, 73}

\bibitem[\protect\citeauthoryear{{Langer}}{{Langer}}{2012}]{Langer_2012}
{Langer} N.,  2012, \mn@doi [\araa] {10.1146/annurev-astro-081811-125534},
  \href {http://adsabs.harvard.edu/abs/2012ARA26A..50..107L} {50, 107}

\bibitem[\protect\citeauthoryear{Lin, Shrestha, Wolfe, Hoffman  \& Stencel}{Lin
  et~al.}{2019}]{Lin_2019}
Lin A.~A.,  Shrestha M.,  Wolfe T.~M.,  Hoffman J.~L.,   Stencel R.~E.,  2019,
  \mn@doi [Research Notes of the {AAS}] {10.3847/2515-5172/ab3d45}, 3, 121

\bibitem[\protect\citeauthoryear{{Martin} \& {Whittet}}{{Martin} \&
  {Whittet}}{1990}]{martin1990}
{Martin} P.~G.,  {Whittet} D.~C.~B.,  1990, \apj, \href
  {https://ui.adsabs.harvard.edu/abs/1990ApJ...357..113M} {357, 113}

\bibitem[\protect\citeauthoryear{{Martin} et~al.,}{{Martin}
  et~al.}{1992}]{martin1992}
{Martin} P.~G.,  et~al., 1992, \apj, \href
  {https://ui.adsabs.harvard.edu/abs/1992ApJ...392..691M} {392, 691}

\bibitem[\protect\citeauthoryear{{Mathis}, {Rumpl}  \& {Nordsieck}}{{Mathis}
  et~al.}{1977}]{Mathis_1977}
{Mathis} J.~S.,  {Rumpl} W.,   {Nordsieck} K.~H.,  1977, \mn@doi [\apj]
  {10.1086/155591}, \href {http://adsabs.harvard.edu/abs/1977ApJ...217..425M}
  {217, 425}

\bibitem[\protect\citeauthoryear{{Meyer}, {Mackey}, {Langer}, {Gvaramadze},
  {Mignone}, {Izzard}  \& {Kaper}}{{Meyer} et~al.}{2014}]{Meyer_2014}
{Meyer} D.~M.-A.,  {Mackey} J.,  {Langer} N.,  {Gvaramadze} V.~V.,  {Mignone}
  A.,  {Izzard} R.~G.,   {Kaper} L.,  2014, \mn@doi [\mnras]
  {10.1093/mnras/stu1629}, \href
  {http://adsabs.harvard.edu/abs/2014MNRAS.444.2754M} {444, 2754}

\bibitem[\protect\citeauthoryear{{Meyer}, {Langer}, {Mackey}, {Vel{\'a}zquez}
  \& {Gusdorf}}{{Meyer} et~al.}{2015}]{Meyer_2015}
{Meyer} D.~M.-A.,  {Langer} N.,  {Mackey} J.,  {Vel{\'a}zquez} P.~F.,
  {Gusdorf} A.,  2015, \mn@doi [\mnras] {10.1093/mnras/stv898}, \href
  {http://adsabs.harvard.edu/abs/2015MNRAS.450.3080M} {450, 3080}

\bibitem[\protect\citeauthoryear{{Meyer}, {Mignone}, {Kuiper}, {Raga}  \&
  {Kley}}{{Meyer} et~al.}{2017}]{Meyer_2017}
{Meyer} D.~M.~A.,  {Mignone} A.,  {Kuiper} R.,  {Raga} A.~C.,   {Kley} W.,
  2017, \mn@doi [\mnras] {10.1093/mnras/stw2537}, \href
  {https://ui.adsabs.harvard.edu/abs/2017MNRAS.464.3229M} {464, 3229}

\bibitem[\protect\citeauthoryear{{Mohamed}, {Mackey}  \& {Langer}}{{Mohamed}
  et~al.}{2012}]{Mohamed_2012}
{Mohamed} S.,  {Mackey} J.,   {Langer} N.,  2012, \mn@doi [\aap]
  {10.1051/0004-6361/201118002}, \href
  {http://adsabs.harvard.edu/abs/2012A26A...541A...1M} {541, A1}

\bibitem[\protect\citeauthoryear{{Moultaka}, {Eckart}, {Viehmann}, {Mouawad},
  {Straubmeier}, {Ott}  \& {Sch{\"o}del}}{{Moultaka}
  et~al.}{2004}]{moultaka2004}
{Moultaka} J.,  {Eckart} A.,  {Viehmann} T.,  {Mouawad} N.,  {Straubmeier} C.,
  {Ott} T.,   {Sch{\"o}del} R.,  2004, \aap, \href
  {https://ui.adsabs.harvard.edu/abs/2004A&A...425..529M} {425, 529}

\bibitem[\protect\citeauthoryear{{Neilson}, {Ignace}, {Shrestha}, {Hoffman}  \&
  {Mackey}}{{Neilson} et~al.}{2013}]{Neilson_2013}
{Neilson} H.~R.,  {Ignace} R.,  {Shrestha} M.,  {Hoffman} J.~L.,   {Mackey} J.,
   2013, in Massive Stars: From alpha to Omega. p.~172

\bibitem[\protect\citeauthoryear{{Neilson}, {Ignace}, {Smith}, {Henson}  \&
  {Adams}}{{Neilson} et~al.}{2014}]{Neilson_2014}
{Neilson} H.~R.,  {Ignace} R.,  {Smith} B.~J.,  {Henson} G.,   {Adams} A.~M.,
  2014, \mn@doi [\aap] {10.1051/0004-6361/201424037}, \href
  {http://adsabs.harvard.edu/abs/2014A26A...568A..88N} {568, A88}

\bibitem[\protect\citeauthoryear{{Quirrenbach} et~al.,}{{Quirrenbach}
  et~al.}{1997}]{quirrenbach1997}
{Quirrenbach} A.,  et~al., 1997, \mn@doi [\apj] {10.1086/303854}, \href
  {https://ui.adsabs.harvard.edu/abs/1997ApJ...479..477Q} {479, 477}

\bibitem[\protect\citeauthoryear{{Rauch} et~al.,}{{Rauch}
  et~al.}{2013}]{Rauch_2013}
{Rauch} C.,  et~al., 2013, \mn@doi [\aap] {10.1051/0004-6361/201219874}, \href
  {http://adsabs.harvard.edu/abs/2013A26A...551A..35R} {551, A35}

\bibitem[\protect\citeauthoryear{{Sanchez-Bermudez}, {Sch{\"o}del}, {Alberdi},
  {Muzi{\'c}}, {Hummel}  \& {Pott}}{{Sanchez-Bermudez}
  et~al.}{2014}]{Sanchez-Bermudez_2014}
{Sanchez-Bermudez} J.,  {Sch{\"o}del} R.,  {Alberdi} A.,  {Muzi{\'c}} K.,
  {Hummel} C.~A.,   {Pott} J.-U.,  2014, \mn@doi [\aap]
  {10.1051/0004-6361/201423657}, \href
  {http://adsabs.harvard.edu/abs/2014A26A...567A..21S} {567, A21}

\bibitem[\protect\citeauthoryear{{Scoville} \& {Burkert}}{{Scoville} \&
  {Burkert}}{2013}]{Scoville_2013}
{Scoville} N.,  {Burkert} A.,  2013, \mn@doi [\apj]
  {10.1088/0004-637X/768/2/108}, \href
  {https://ui.adsabs.harvard.edu/abs/2013ApJ...768..108S} {768, 108}

\bibitem[\protect\citeauthoryear{{Serkowski}, {Mathewson}  \&
  {Ford}}{{Serkowski} et~al.}{1975}]{Serkowski_1975}
{Serkowski} K.,  {Mathewson} D.~S.,   {Ford} V.~L.,  1975, \mn@doi [\apj]
  {10.1086/153410}, \href {http://adsabs.harvard.edu/abs/1975ApJ...196..261S}
  {196, 261}

\bibitem[\protect\citeauthoryear{{Shahzamanian} et~al.,}{{Shahzamanian}
  et~al.}{2016}]{Shahzamanian_2016}
{Shahzamanian} B.,  et~al., 2016, \mn@doi [\aap] {10.1051/0004-6361/201628994},
  \href {http://adsabs.harvard.edu/abs/2016A26A...593A.131S} {593, A131}

\bibitem[\protect\citeauthoryear{{Shrestha}, {Neilson}, {Hoffman}  \&
  {Ignace}}{{Shrestha} et~al.}{2018}]{Shrestha_2018}
{Shrestha} M.,  {Neilson} H.~R.,  {Hoffman} J.~L.,   {Ignace} R.,  2018,
  \mn@doi [\mnras] {10.1093/mnras/sty724}, \href
  {http://adsabs.harvard.edu/abs/2018MNRAS.tmp..711S} {}

\bibitem[\protect\citeauthoryear{{Smith}, {Hinkle}  \& {Ryde}}{{Smith}
  et~al.}{2009}]{Smith_2009}
{Smith} N.,  {Hinkle} K.~H.,   {Ryde} N.,  2009, \mn@doi [\aj]
  {10.1088/0004-6256/137/3/3558}, \href
  {http://adsabs.harvard.edu/abs/2009AJ....137.3558S} {137, 3558}

\bibitem[\protect\citeauthoryear{{Tanner}, {Ghez}, {Morris}, {Becklin},
  {Cotera}, {Ressler}, {Werner}  \& {Wizinowich}}{{Tanner}
  et~al.}{2002}]{Tanner_2002}
{Tanner} A.,  {Ghez} A.~M.,  {Morris} M.,  {Becklin} E.~E.,  {Cotera} A.,
  {Ressler} M.,  {Werner} M.,   {Wizinowich} P.,  2002, \mn@doi [\apj]
  {10.1086/341470}, \href
  {https://ui.adsabs.harvard.edu/abs/2002ApJ...575..860T} {575, 860}

\bibitem[\protect\citeauthoryear{{Ueta} et~al.,}{{Ueta}
  et~al.}{2008}]{Ueta_2008}
{Ueta} T.,  et~al., 2008, \mn@doi [\pasj] {10.1093/pasj/60.sp2.S407}, \href
  {http://adsabs.harvard.edu/abs/2008PASJ...60S.407U} {60, S407}

\bibitem[\protect\citeauthoryear{{Whitney}}{{Whitney}}{2011}]{whitney_2011}
{Whitney} B.~A.,  2011, Bulletin of the Astronomical Society of India, \href
  {http://adsabs.harvard.edu/abs/2011BASI...39..101W} {39, 101}

\bibitem[\protect\citeauthoryear{Whitney \& Wolff}{Whitney \&
  Wolff}{2002}]{Whitney_2002}
Whitney B.~A.,  Wolff M.~J.,  2002, \apj, 574, 205

\bibitem[\protect\citeauthoryear{{Whitney}, {Robitaille}, {Bjorkman}, {Dong},
  {Wolff}, {Wood}  \& {Honor}}{{Whitney} et~al.}{2013}]{whitney_2013}
{Whitney} B.~A.,  {Robitaille} T.~P.,  {Bjorkman} J.~E.,  {Dong} R.,  {Wolff}
  M.~J.,  {Wood} K.,   {Honor} J.,  2013, \mn@doi [\apjs]
  {10.1088/0067-0049/207/2/30}, \href
  {http://adsabs.harvard.edu/abs/2013ApJS..207...30W} {207, 30}

\bibitem[\protect\citeauthoryear{{Wilkin}}{{Wilkin}}{1996}]{Wilkin_1996}
{Wilkin} F.~P.,  1996, \mn@doi [\apjl] {10.1086/309939}, \href
  {http://adsabs.harvard.edu/abs/1996ApJ...459L..31W} {459, L31}

\bibitem[\protect\citeauthoryear{{Wood}, {Bjorkman}, {Whitney}  \&
  {Code}}{{Wood} et~al.}{1996}]{Wood_96a}
{Wood} K.,  {Bjorkman} J.~E.,  {Whitney} B.~A.,   {Code} A.~D.,  1996, \mn@doi
  [\apj] {10.1086/177105}, \href
  {http://adsabs.harvard.edu/abs/1996ApJ...461..828W} {461, 828}

\bibitem[\protect\citeauthoryear{{Zaja{\v{c}}ek} et~al.,}{{Zaja{\v{c}}ek}
  et~al.}{2017a}]{zajacksum_2017}
{Zaja{\v{c}}ek} M.,  et~al., 2017a, in RAGtime 17-19: Workshops on Black Holes
  and Neutron Stars. pp 237--252

\bibitem[\protect\citeauthoryear{{Zaja{\v{c}}ek} et~al.,}{{Zaja{\v{c}}ek}
  et~al.}{2017b}]{Zajacek_2017}
{Zaja{\v{c}}ek} M.,  et~al., 2017b, \mn@doi [\aap]
  {10.1051/0004-6361/201730532}, \href
  {https://ui.adsabs.harvard.edu/abs/2017A&A...602A.121Z} {602, A121}

\bibitem[\protect\citeauthoryear{{Zubko} \& {Laor}}{{Zubko} \&
  {Laor}}{2000}]{Zubko_2001}
{Zubko} V.~G.,  {Laor} A.,  2000, \mn@doi [\apjs] {10.1086/313373}, \href
  {https://ui.adsabs.harvard.edu/abs/2000ApJS..128..245Z} {128, 245}

\bibitem[\protect\citeauthoryear{{van Marle}, {Meliani}, {Keppens}  \&
  {Decin}}{{van Marle} et~al.}{2011}]{Marle_2011}
{van Marle} A.~J.,  {Meliani} Z.,  {Keppens} R.,   {Decin} L.,  2011, \mn@doi
  [\apjl] {10.1088/2041-8205/734/2/L26}, \href
  {https://ui.adsabs.harvard.edu/abs/2011ApJ...734L..26V} {734, L26}

\bibitem[\protect\citeauthoryear{{van Marle}, {Decin}, {Cox}  \&
  {Meliani}}{{van Marle} et~al.}{2015}]{Marle_2015}
{van Marle} A.~J.,  {Decin} L.,  {Cox} N.~L.~J.,   {Meliani} Z.,  2015, in
  Journal of Physics Conference Series. p. 012024,
  \mn@doi{10.1088/1742-6596/577/1/012024}

\makeatother
\end{thebibliography}







\end{document}